\global\let\ifmypprint\iffalse 
\def\mypprint{\global\let\ifmypprint\iftrue}
\global\let\iftorefs\iffalse
\def\torefs{\global\let\iftorefs\iftrue}
\global\let\dofloatfig\iffalse
\def\floatthefig{\let\dofloatfig\iftrue}

\mypprint 
\ifmypprint
\documentstyle[twocolumn,eqsecnum,prl,aps]{revtex}
\floatthefig
\else
  \iftorefs 
    \catcode`\@=11 
    
    \def\figure{\let\@capwidth\columnwidth\@float{figure}}
    \let\endfigure\end@float
    \@namedef{figure*}{\let\@capwidth\textwidth\@dblfloat{figure}}
    \@namedef{endfigure*}{\end@dblfloat}
    \catcode`\@=12 
    \floatthefig 
  \fi
\fi

\input epsf.tex

\begin{document}

\twocolumn[\hsize\textwidth\columnwidth\hsize\csname @twocolumnfalse\endcsname

\title{The Shapes of Flux Domains in the Intermediate State \\ of 
Type-I Superconductors}

\author{Alan T. Dorsey\cite{emaild}}
\address{Department of Physics, University of Florida, Gainesville, FL  32611}

\author{Raymond~E. Goldstein\cite{emailr}} 
\address{Department of Physics and Program in Applied Mathematics}
\address{University of Arizona, Tucson, AZ  85721}

\date{\today}

\maketitle

\begin{abstract}

In the intermediate state of a thin type-I superconductor magnetic flux
penetrates in a disordered set of highly branched and fingered
macroscopic domains.  To understand these shapes, we study in detail 
a recently proposed ``current-loop" (CL) model {[}R.E. Goldstein, 
D.P. Jackson, A.T. Dorsey, {\it Phys. Rev. Lett.} {\bf 76}, 3818 (1996){]} 
that models the intermediate state as a collection of tense current ribbons
flowing along the superconducting-normal interfaces and subject to the 
constraint of global flux conservation.  
The validity of this model is tested through
a detailed reanalysis of Landau's original conformal mapping treatment of the 
laminar state, in which the superconductor-normal interfaces are
flared within the slab, and of a closely-related straight-lamina model.
A simplified dynamical model is described that elucidates the nature of 
possible shape instabilities of flux stripes and stripe arrays, 
and numerical studies of the highly nonlinear regime of those 
instabilities demonstrate patterns like those seen experimentally.
Of particular interest is the buckling instability commonly seen in 
the intermediate state.  The free-boundary approach further allows for
a calculation of the elastic properties of the laminar state, which
closely resembles that of smectic liquid crystals.
We suggest several new experiments to 
explore of flux domain shape instabilities, including
an Eckhaus instability induced by changing the out-of-plane magnetic field,
and an analog of the Helfrich-Hurault instability of smectics 
induced by an in-plane field.

\end{abstract}
\vskip2pc]

\section{Introduction}

\begin{figure}
\epsfxsize=2.8 truein
\centerline{\epsffile{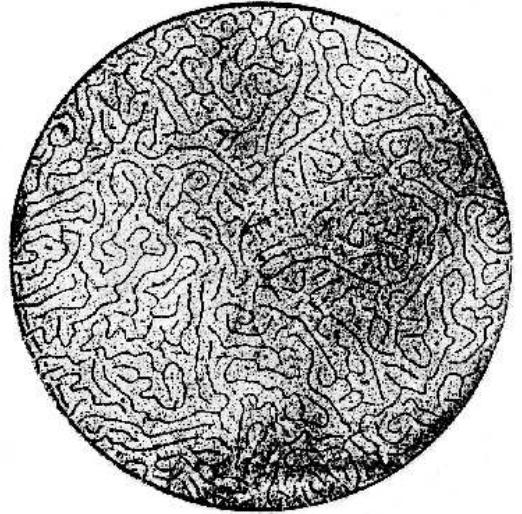}}
\smallskip
\caption[]{The intermediate state of a thin slab of indium, in which
the superconducting regions (black) are decorated with
niobium (black). The applied
field $H_a$ is close to the critical field $H_c$ ($h=H_a/H_c=0.931$).
Adapted from Haenssler and Rinderer \protect{\cite{Haenssler}}.}
\label{expt_fig}
\end{figure}

A longstanding problem in macroscopic superconductivity
is that of understanding the complex patterns of
flux penetration observed in the intermediate state of a type-I
superconductor. This state is observed when a
thin superconducting slab is placed in a perpendicular magnetic field.  
Unlike type-II superconductors, where the field penetration is in the form
of tubes each with a quantum of magnetic flux, type-I systems are
observed to form intricately branched and fingered {\it macroscopic} 
flux domains \cite{Huebener,Faber,Haenssler}.  Thus, instead of
establishing a Meissner phase, in which the magnetic induction ${\bf B}=0$
uniformly, the demagnetizing effects of the large aspect ratio force the
sample to break up into regions, some of which are
uniformly superconducting (with ${\bf B=0}$ inside) and others that
are normal (${\bf B}\ne 0$).  Figure \ref{expt_fig} shows 
a typical example of these patterns \cite{Haenssler}.  The superconducting
regions appear black, having been decorated with a powder (niobium) that
is itself superconducting at the sample temperature and thus migrates 
to the regions of low magnetic field.  
Other imaging techniques include Hall probes \cite{Goren} and 
magneto-optics \cite{magnetooptic}.  All reveal similar structures.

The sample in Fig.~\ref{expt_fig} is
at an applied magnetic field $H_a$ that is very close to the critical
field $H_c$ at which the sample would be completely normal, so the
minority phase is superconducting.  Similar patterns are observed
when $H_a/H_c$ is very small, but now the minority
phase
\dofloatfig
\fi
is normal; the sample consists of fingered and branched 
{\it flux domains} in a
matrix of superconductor.  These domains have
a characteristic field-dependent finger width, and the branched domains
have three-fold vertices.

For many years it has been known \cite{Sharvin} that more ordered
flux domain states may be observed in the presence of a small in-plane
component of the applied magnetic field.  It is also known that
the domain morphology is not a thermodynamic state function; 
it depends on the path in field-temperature space through
which the sample has been brought to a given point \cite{Haenssler}.
Thus, for instance, cooling in zero field below the transition
temperature and then applying the field tends to produce patterns
in which normal domains are embedded in a matrix of superconductor,
whereas when the same point in $T-H$ space is reached by cooling below
$T_c$ in a fixed field the normal domains connect to the sample edges
\cite{Huebener}.
These observations suggest that the patterns are not
true ground states of the system---the sample is not in true
thermodynamic equilibrium.

\begin{figure}
\epsfxsize=3.3 truein
\centerline{\epsffile{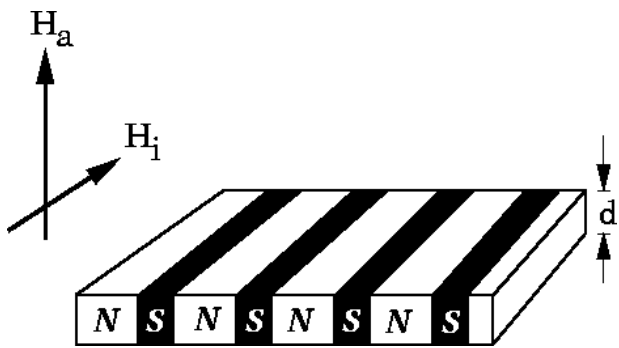}}
\smallskip
\caption[]{The laminar state in a thin slab.  The applied field $H_a$ is
normal to the slab.  An ordered laminar structure is obtained with an
additional in-plane component $H_i$.}
\label{laminar_fig}
\end{figure}

Despite the ubiquity of these patterns, there has until recently
been no theoretical explanation for their form.  The earliest
attempt, prior to the detailed experiments described above, was
by Landau \cite{Landau}, who considered the laminar state: a periodic array of
alternating superconducting and normal domains (Fig.~\ref{laminar_fig}). 
Exploiting the translational invariance
of the pattern in the direction parallel to the stripes, the
cross-sectional shape of the domain walls and the associated bending
of the magnetic field lines become purely two-dimensional problems
amenable to conformal mapping techniques.  The primary result of this
calculation is a determination of the laminar state free energy as a function
of lamina spacing, applied field, superconductor-normal surface energy, 
and slab thickness.

\dofloatfig
\fi

The free energy so obtained depends on two parameters:
the area fraction occupied by the normal state and
the repeat distance of the structure.  For thick slabs,
one finds to a good approximation that the equilibrium
area fraction is set by the reduced field $h=H_a/H_c$.
Deep within the slab the domain
walls are aligned with the field, but they flare along with
the magnetic field lines near the surface.  
The equilibrium field-dependent width has been found to be in good agreement 
with subsequent experiments \cite{Huebener,Faber,Haenssler}.
Stability calculations about this state are, however, precluded by
the reliance on conformal mapping techniques, and no systematic
calculations of this type have been performed.
Subsequently, and in light of experiments showing nearly circular
flux domains, there have been several calculations of the energies of
periodic arrays of simple geometric structures \cite{others}.
These studies generally find that the free energies of these periodic
structures are very close to each other, so there is little thermodynamic
driving force preferring one over the other.
None of these theories offers an explanation for the appearance of
branched flux domains.

The dual features of (i) disordered domain shapes and (ii) path dependence to the
patterns makes it clear that an understanding of the intermediate state
requires a theoretical approach that (a) treats the superconductor-normal
(SN) interfaces as a {\it free-boundary} problem, 
and (b) addresses the {\it dynamics}
of those interfaces.  That one should consider
an interfacial representation at all, rather than a more microscopic
approach based on equations of motion for the superconducting order parameter
and the vector potential, is made clear by the strong {\it separation of scales}
that exists between the domain size (typically fractions of a millimeter)
and the width of the SN interface.  For a strongly type-I superconductor
the width is set by the coherence length $\xi$, which is on the order
of $0.2$ $\mu$m.

Several recent studies \cite{Dorsey91,Liu91} have
emphasized strong connections between the motion of SN
interfaces in the presence of a magnetic field and the dynamics of
solid-liquid interfaces in the process of solidification.
The key to this relationship is the fact that the magnetic field
in the normal state obeys a diffusion equation \cite{Pippard}, analogous 
to the diffusion of latent heat in the solidification problem.  
These suggest that a {\it diffusional instability} like the Mullins-Sekerka
instability \cite{Mullins} should occur in the process of flux invasion.
Numerical studies of the time-dependent Ginzburg-Landau model 
confirmed these instabilities, which can lead to highly ramified domain
patterns.  

It was also found by asymptotic methods \cite{asymptotics}
that in the limit of a sharp interface it is possible to integrate out
the magnetic field contributions and arrive at a nonlocal free-boundary
theory for the SN interface alone.  The nonlocality is both temporal
and spatial, the latter taking the form of a 
Biot-Savart interaction between distant segments of the interface.
The appearance of this type of coupling reminds us that the supercurrents
that provide the screening of the applied magnetic field flow along
the SN interfaces.  
Many years ago
Pearl \cite{Pearl}, and later Fetter and Hohenberg \cite{Fetter} 
showed that the interactions between currents flowing in thin slabs
are long-ranged.  Thus, vortices in a thin film interact with a potential
whose long-range form is unscreened, $V(r)\sim 1/r$, while at short
distances $V(r)\sim \ln(\Lambda/r)$, where $\Lambda$ is a cutoff.
Screening is unimportant at long distances because the interaction energy
is dominated by the electromagnetic fields ({\it in vacuo})
above and below the slab.

Based on all of these observations, we recently proposed
\cite{goldstein96} a ``current-loop" (CL) model as an approximate
description of the intermediate state.  The model describes the
patterns as a collection of tense ribbons of current, interacting with
the long-range Biot-Savart interaction of currents in free space, and
subject to a constraint of global flux conservation.  It is based on
several simplifying assumptions.  First, the domain walls are taken to
be vertical, rather than flared as in Landau's more precise calculation.
Second, only the supercurrents in those walls are considered---surface
supercurrents on the top and bottom of the sample are ignored.
These approximations allow for an averaging process over the slab thickness
that maps the simplified model onto one of self-interacting contours 
in the plane.
Such free-boundary approaches are well-known for hydrodynamic
problems such as Saffman-Taylor fingering \cite{Kesslerrev}, as well as in
systems governed by partial differential equations of the 
reaction-diffusion type.  In the latter case, the dynamics may be
reduced to that of an interface when the scale of the patterns is
large compared to the width of the front connecting two
locally stable states.  

While several important phenomena are precluded from study in this model
(such as domain fission), it has the virtue of providing a simple picture of
the mechanism of shape instabilities in this system, and may form a useful
starting point for a more precise treatment \cite{Narayan}.
It also is strikingly similar to models for the energetics and dynamics of
interfacial pattern formation in a variety of other systems that display
``labyrinthine" patterns from a competition between interfacial tension
and long-range electromagnetic interactions \cite{Seulscience}.  These include
fingering instabilities of magnetic fluids in Hele-Shaw 
flow \cite{Rosensweig,cebers,science,pra,pre}, 
Langmuir monolayers at the air-water interface 
\cite{Andelman,LeeMcConnell,jpc} and thin magnetic films \cite{Seulfilms}.
It is also equivalent to a reaction-diffusion system 
\cite{turingprl,turingpre,Hagberg,Muratov}
in which {\it chemical fronts} between two locally stable states may form
space-filling disordered labyrinths similar to those seen in experiments
on chemical pattern formation in a gel reactor \cite{Lee,Leelong}.

In Section II we set the stage for a macroscopic model of the
intermediate state by reviewing both the sharp-interface limit of
the time-dependent Ginzburg-Landau model for nonequilibrium 
superconductivity and the derivation of long-range interactions
between currents in the slab geometry.
The conformal mapping solution to Landau's model of the intermediate
state and the simpler {\it straight-lamina} model are derived in
Section III, in which careful attention is paid to the consequences of
those long-range interactions.  The structure of the free energy as
a function of stripe periodicity and width, surface tension and
applied field in the two models is shown to be essentially equivalent, 
as are their predictions for the equilibrium stripe width as a function of
field.  The current-loop model is described and applied to the
energetics of the laminar state in Section IV. Instabilities of single flux
stripes and the elastic properties of the laminar state are found
in Section V.  The concluding Section VI summarizes the
new experimental predictions that arise from the correspondence between
the intermediate state and others such as Langmuir monolayers 
and smectic liquid crystals.
An appendix gathers together calculational details of the
stability analyses and elasticity calculations.

\section{Properties of the macroscopic model} 

\subsection{The sharp-interface limit}

Here we review the results of asymptotic methods applied
to the time-dependent Ginzburg-Landau (TDGL) model of superconductivity
in the limit of sharp interfaces. 
The microscopic parameters that enter the TDGL include 
the charge $e^*$ and mass $m$ of a Cooper pair,
a dimensionless order parameter relaxation time $\gamma$,
and the conductivity $\sigma$ of the normal phase.  The
coupled equations of motion for the order parameter $\psi$ and
the scalar and magnetic vector potentials,
$\phi$ and ${\bf A}$, are 
\begin{eqnarray}
\hbar\gamma\left(\partial_t + {ie^*\over \hbar}\phi\right)\psi
&=&{\hbar^2\over 2m}\left(\bbox{\nabla}
-{ie^*\over \hbar}{\bf A}\right)^{\!2}\psi \\
&&+ a\psi-b \vert\psi\vert^2\psi, \nonumber \\
\bbox{\nabla}\times \bbox{\nabla}\times {\bf A}&=&
4\pi\left({\bf J}_n+{\bf J}_s\right)~.
\label{tdgl}
\end{eqnarray}
where ${\bf J}_n$ and ${\bf J}_s$ are the normal and supercurrents,
\begin{eqnarray}
{\bf J}_n&=&\sigma\left(-\nabla\phi-\partial_t{\bf A}\right)\nonumber \\
{\bf J}_s&=& {\hbar e^*\over 2mi}
\left(\psi^*\bbox{\nabla} \psi-\psi\bbox{\nabla}\psi^*\right)-{{e^*}^2\over m}
\vert\psi\vert^2 {\bf A}~.
\label{currents}
\end{eqnarray}
The bifurcation parameter $a$ contains the important temperature dependence,
$a=a_0\left(1-T/T_c\right)$ and controls the correlation length
$\xi=\hbar/(2m\vert a\vert)^{1/2}$ and penetration depth
$\lambda=\left[mb/4\pi {e^*}^2\vert a \vert \right]^{1/2}$.
Finally, the Ginzburg-Landau parameter $\kappa=\lambda/\xi$.

When $\xi$ is small the superconductor-normal interface will be sharp,
and a natural measure of smallness is provided by the inverse distance from
the critical point.  Setting $a=\tilde a/\epsilon^2$ and rescaling
all distances and fields with appropriate powers of $\epsilon$, one finds
that both $\lambda$ and $\xi$ scale identically.  The sharp-interface
limit is then achieved by $\xi, \lambda\to 0$ with $\kappa$ fixed.
It is thus possible in this limit to continue to distinguish between type-I 
($\kappa<1/\sqrt{2}$) and type-II ($\kappa>1/\sqrt{2}$) systems.

Existing derivations of interface equations of motion in this context
presume translational invariance in the direction along which the magnetic
field is applied, and hence describe ``bulk" superconductors \cite{asymptotics}.
The results may be summarized as follows.  Far away from the interface
(the ``outer" solution in the sense of matched asymptotics), the
magnetic field in the normal region obeys the diffusion equation
\begin{equation}
{\bf h}_t=D\nabla^2 {\bf h}~,
\label{mag_diff_outer}
\end{equation}
with diffusion constant $D=1/4\pi\sigma$.
The boundary condition on the magnitude $h_i$ of ${\bf h}$ on the interface 
${\cal C}$ is
\begin{equation}
h\bigl|_{\cal C}=H_c\left[1-{4\pi\over H_c^2}\left(\sigma_{SN}{\cal K} +
\Gamma^{-1}v_n\right)\right]~.
\label{mag_bc}
\end{equation}
where $\sigma_{SN}$ is the interfacial tension (with dimensions of 
energy/area), ${\cal K}$ is the
interface curvature, $\Gamma$ is a known kinetic coefficient,
and $v_n=\hat{\bf n}\cdot {\bf r}_t$ is the normal component of the 
interface velocity.

The equation of motion is an expression for $v_n$ in terms of
the function ${\bf r}(s,t)$, and it follows from a solution to the 
diffusion equation (\ref{mag_diff_outer}) given the boundary conditions
(\ref{mag_bc}).  As with all such problems, this
involves a convolution over previous times and all space.  But in the
limit $D\to \infty$, or vanishing normal state conductivity, the
temporal nonlocality disappears.  This yields the equation of motion
\begin{eqnarray}
\Gamma^{-1}\hat{\bf n}\cdot {\bf r}_t(s,t)&=&
{H_c^2-H_a^2\over 8\pi}+\sigma_{SN}{\cal K}\nonumber \\
&&-{H_c^2\over 8\pi^2}\oint\! ds'{{\bf R}\times \hat{\bf t}(s')\over R^2}~,
\label{super_eom}
\end{eqnarray}
with ${\bf R}={\bf r}(s)-{\bf r}(s')$.
The dynamics (\ref{super_eom}) has the variational form
\begin{equation}
{\bf r}_t=-\Gamma {1\over \sqrt{g}}{\delta {\cal H}_{\rm eff}
\over \delta {\bf r}}
\label{super_var}
\end{equation}
where $g$ is the interface metric and the effective interface 
Hamiltonian for a single domain is
\begin{eqnarray}
{\cal H}_{\rm eff}[{\bf r}]&=&-{H_c^2-H_a^2\over 8\pi} A
+\sigma_{SN}L\nonumber \\
&&-{H_c^2\over 8\pi}\oint\! ds\oint\! ds'\hat{\bf t}(s)\cdot
\hat{\bf t}(s'){\cal G}(R)~,
\label{super_Ham}
\end{eqnarray}
with ${\cal G}(R) = -(1/2\pi)\ln R$ is the Green's function of the Laplacian in
two dimensions, and $A$ and $L$ the area and perimeter of the domain. 
As anticipated in the introduction, we see that the energy
associated with the distortion of the magnetic field lines
threading a domain is represented by the self-induction of the
boundary.  The other contributions are simply the line energy $\sigma_{SN}L$
and an area term associated with the magnetic pressure and condensation
energy. 

The limits of a sharp interface and zero normal state conductivity
that produce a temporally local but spatially nonlocal contour dynamics
have parallels in a simpler reaction-diffusion system of recent
interest \cite{turingprl,turingpre,Hagberg,Muratov}.  This is the
FitzHugh-Nagumo model \cite{FitzHugh} of the coupled dynamics of 
an activator $u$ and inhibitor $v$, considered in two spatial dimensions,
\begin{eqnarray}
u_t &=& \bar{D}\nabla^2 u - F'(u) - \rho \left(v-u\right)\nonumber \\
\epsilon v_t &=& \nabla^2 v-v+u~.
\label{FN}
\end{eqnarray}
These partial differential equations are written in a rescaled form in which
the activator diffusion constant $\bar{D}$ is normalized to that of
the inhibitor, while $\epsilon$ is a ratio of their characteristic times.
The function $F(u)$, whose derivative $F'(u)$ appears above, 
is a double-well potential that describes the auto-catalytic behavior 
(and bistability) of the activator. 
When $\epsilon\ne 0$, the coupled dynamics is not a gradient flow
in any standard form. 

It is clear that the two systems (\ref{tdgl}) and (\ref{FN}) share
many features.  In each case there is an order parameter field 
($\psi$ or $u$) with
an underlying bifurcation (the former continuous, the latter of first order),
which is coupled to a diffusing field (unscreened or
screened).  In the case of the superconductor, the second field
could be integrated out of the problem in exchange for an instantaneous nonlocal
coupling of the field $\psi$.
An identical feature appears in the reaction-diffusion problem when the
parameter $\epsilon$ is small, for then the inhibitor relaxes on times short 
compared to that of the activator and will be slaved to $u$.  
Setting $\epsilon v_t=0$,
the field $v$ may then be expressed as $v=\int {\cal G} u$, where ${\cal G}$
is the appropriate Green's function for the modified Helmholtz operator.
This produces the nonlocal dynamics for $u$ that {\it is} variational,
$u_t = -\delta {\cal E}/\delta u$, with
\begin{eqnarray}
{\cal E}[u]&=&\int\! d^2x
\left\{{1\over 2} \bar{D} \vert\bbox{\nabla} u\vert^2 
+ F(u)-{1\over 2}\rho^2\right\}\nonumber \\
&&+{1\over 2}\rho \int\! d^2x\!\int\! d^2x' 
u({\bf x}){\cal G}(|{\bf x}-{\bf x}'|) u({\bf x}')~.
\label{slaved_energy}
\end{eqnarray}

Now taking the limit $\bar{D}\to 0$ we may
reduce (\ref{slaved_energy}) to a
functional of the contours bounding regions in which the
activator takes on values corresponding to different minima of $F$.
That new functional (of the position vectors ${\bf r}_i(s)$ for each
domain boundary, parameterized by arclength $s$) is 
\cite{turingprl,turingpre}
\begin{eqnarray}
\Delta{\cal E}[\left\{{\bf r}_i\right\}] &=&\bar{\gamma} \sum_i L_i + 
\Delta F \sum_i A_i \nonumber \\
&&-{\rho\over 2}\sum_{i,j}\oint\! ds\oint\! ds'
\hat{\bf t}_i \cdot \hat{\bf t}_j
{\cal G}\left(|{\bf r}_i-{\bf r}_j|\right),
\label{energyfunc}
\end{eqnarray}
with $\bar{\gamma}$ the line tension, and where $L_i$ and $A_i$ the 
perimeter and enclosed area of each domain.
The nonlocal term is again in the form of the self- and mutual-induction
of (here fictitious) current loops encircling each of the domains.

\subsection{Long-range forces}

As further motivation for the CL model, we recall Pearl's derivation \cite{Pearl}
of the long-range (unscreened) potential between point vortices 
in a thin film.  The starting point is the solution for the
supercurrent surrounding a single vortex located at the origin.
In the slab geometry, the order parameter has the simple form
\begin{equation}
\psi({\bf r})=\psi_0f(r)e^{i\theta} \ \ \ \ \left(0\le z \le d\right)
\label{psi_slab}
\end{equation}
where $\psi_0$ is the far-field value of $\psi$, which vanishes 
for $z<0$ and $z> d$. The function $f(r)$ describes the
vortex core structure, and we take $f=1$ on the (large) scales of interest.
The second of the Ginzburg-Landau pair of equations (\ref{tdgl})
is then
\begin{equation}
\bbox{\nabla}\times \bbox{\nabla}\times {\bf A}=
4\pi{\bf J}_s={1\over \lambda^2}\left[{\psi_0\over 2\pi r}
\hat{\bf \theta}-{\bf A}\right]~.
\label{tdgl2_pearl}
\end{equation}
If the film is sufficiently thin, then we may average over the film
thickness $d$ to obtain
\begin{equation}
\bbox{\nabla}\times \bbox{\nabla}\times {\bf A}=
{d\over \lambda^2}\left[{\psi_0\over 2\pi r}
\hat{\bf \theta}-{\bf A}\right]\delta(z)~.
\label{tdgl2_pearl2}
\end{equation}
In this form, it is readily apparent from the appearance of
the term $-(d{\bf A}/\lambda^2)\delta(z)$ on the right-hand-side 
that the screening is confined solely to the slab.

The solution to this can be obtained through the use of
Fourier-Hankel transforms and yields the supercurrent
\begin{equation}
{\bf J}_s(r) = -{\phi_0\over 8\pi \Lambda^2}\left[H_1(r/\Lambda)
-Y_1(r/\Lambda)-{2\over \pi}\right]\hat{\bf \theta}
\label{tdgl2_pearl3}
\end{equation}
where $H_1$ and $Y_1$ are Hankel functions, and  the effective penetration
depth for the thin film is 
\begin{equation}
\Lambda={2\lambda^2\over d}~.
\label{Lambda_define}
\end{equation}
The interaction potential of a second point vortex placed a distance
$r$ from the first is obtained by multiplying (\ref{tdgl2_pearl3}) by
$\phi_0$ and integrating, with the result
\begin{equation}
V(r)={\phi_0^2\over 8\pi\Lambda}\left[H_0\left(r/\Lambda\right)
-Y_0\left(r/\Lambda\right)\right]~.
\label{pearl_interaction}
\end{equation}
At long distances $r/\Lambda\gg 1$, this is an unscreened potential,
\begin{equation}
V(r)\simeq {\phi_0^2\over 4\pi^2 r}~,
\label{pearl_long}
\end{equation}
whereas as $r\to 0$ the familiar logarithmic interaction between
vortices appears,
\begin{equation}
V(r)\simeq {\phi_0^2\over 4\pi^2\Lambda}
\log\left( {e^C r\over 2\Lambda}\right)~,
\label{pearl_short}
\end{equation}
with $C$ being Euler's constant.
We anticipate therefore that the interactions between Meissner currents
in a thin type-I slab should have a similar long-range character, and
now proceed to a detailed calculation.

\section{The laminar state} 
\label{laminar_state}

\subsection{Landau's free-boundary solution}

In this first part of our discussion, we calculate the shape and 
optimal spacing for an assumed laminar geometry for the intermediate 
state.  While our work reproduces  Landau's original calculation \cite{Landau}, 
our method is quite different, and we can provide an explicit expression for the 
function $f(h)$ (see below).  We are including this material in our discussion 
since (1) Landau's derivation is, to our taste, a bit obscure, so we hope that 
the present derivation will clarify the techniques and inspire further 
work on extending the calculations to more complicated geometries;  
and (2) the ``exact'' result derived here, for flared normal lamina, 
can be compared against the results of the current loop model discussed
below, allowing us to ``calibrate'' the current loop model. 

\begin{figure}
\epsfxsize=3.3 truein
\centerline{\epsffile{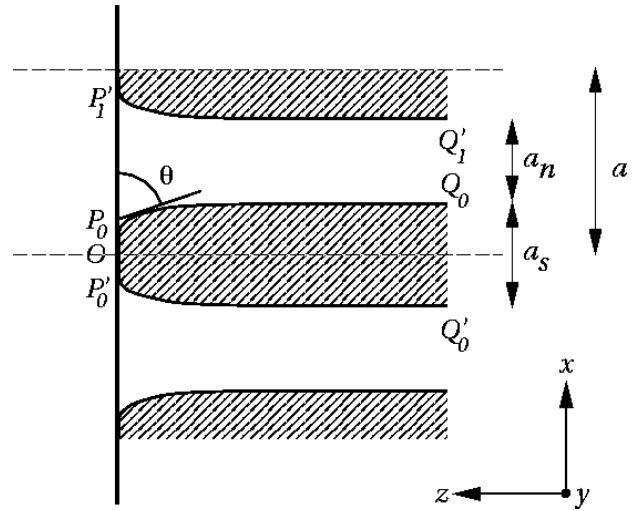}}
\smallskip
\caption[]{Laminae in the intermediate state.  Superconducting regions are
shaded.  Reference points $P_n$, $Q_n$, $P_n'$, etc. are discussed in text.}
\label{diagram}
\end{figure}

We consider the simplest semi-infinite geometry here, which we have 
illustrated in Fig. \ref{diagram}.  
The surface of the material is in the $x-y$ plane, occupying the 
region $z<0$.  The field is applied perpendicular to the sample, so that 
${\bf H}_{a} = H_{a} \hat{z}$.   We will assume that there is an 
array of normal-superconducting laminae periodic in the $x$-direction, 
the laminae being straight in the $y$-direction.   The normal 
and superconducting laminae have asymptotic widths $a_n$ and $a_s$,
respectively, so that the period of the entire structure is 
$a=a_s + a_n$.  

\dofloatfig
\fi

We start by noting that the typical lamina spacing is large compared to the 
superconducting penetration depth and the coherence length, so that the 
interfaces separating the normal and superconducting phase are sharp. 
Then, following Landau, we can work with the macroscopic Maxwell equations. 
In the normal regions we have $\nabla \times {\bf B} = 0$, 
$\nabla \cdot {\bf B} = 0$, which can be solved either by introducing the 
scalar potential $\phi$ through ${\bf B} = - \nabla \phi$, so that 
$\nabla^2 \phi = 0$, or the  vector potential ${\bf A} $ through  
${\bf B} = \nabla \times {\bf A}$, which in the gauge $\nabla\cdot {\bf A}=0$
satisfies $\nabla^2 {\bf A} = 0$.  We thus need to solve Laplace's equation 
in the $x-z$ plane.  This immediately suggests the use of complex 
variable methods; indeed,  for the laminar structure the only 
nonzero component of the vector potential is $A_y$, so we may introduce the 
complex potential 
\begin{equation}
 w = \phi + i \psi= \phi + i A_y 
\label{w}
\end{equation}
so that the complex magnetic field is 
\begin{equation}
B = B_x - i B_z = - {d w \over d\zeta}, \qquad \zeta = x + iz.
\label{H}
\end{equation}
The boundary conditions are the following. First, as $z\rightarrow \infty$
the field becomes the uniform applied field $H_{a} \hat{z}$, so that 
\begin{equation}
w\sim i H_{a} \zeta, \qquad z\rightarrow\infty~.
\label{bc1}
\end{equation}
Since the magnetic field vanishes in the superconducting regions,
continuity of its normal component implies that the field is purely tangential
both along the normal-superconducting interface $P_nQ_n$ and along the segment
which we refer to as the ``nose" ($OP_n$) (Fig.~\ref{diagram}).
In the first case, the assumption of local thermodynamic equilibrium
at this interface implies that the magnitude $H_n$ of the tangential component
is a constant (as yet unknown), so that the vector potential 
$A_y$ is also constant along any one 
interface.  Along the nose 
the magnetic field is parallel to the surface (${\bf H}=H_x\hat x$)
but with a magnitude that is no longer constant.
The field $H_{n}$ along with the 
periods $a_n$ and $a_s$ are determined {\it a posteriori} by 
minimizing the energy.  However, these parameters are not completely 
independent, since by flux conservation we must have $H_a a = H_n a_n$. 
Therefore for fixed external field $H_a$ the energy will be determined
by $a$ and $a_n$.  

\subsubsection{Exact determination of the lamina shape}
 
\begin{figure}
\epsfxsize=3.0 truein
\centerline{\epsffile{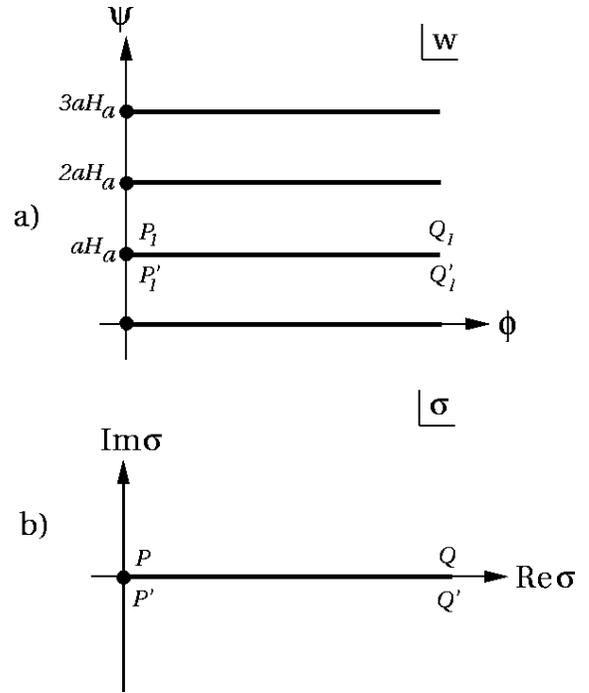}}
\smallskip
\caption[]{The $w$- and $\sigma$-planes.}
\label{wsigmaplane}
\end{figure}

The position of the interface separating the normal and 
superconducting regions is not 
known {\it a priori}, and must be discovered in the process of solving 
the problem.  Although this sounds like a formidable task, it is 
made easier by recognizing that our magnetostatics problem is 
formally equivalent to the flow of an ideal incompressible fluid around
an array of plates, the plates being the noses of the laminae.   The 
field lines would be the streamlines of the fluid; the normal-superconducting
interfaces correspond to  fluid streamlines which have separated from the 
flow behind the plate (``free-streamlines'').   This correspondence is 
outlined in more detail in Table~\ref{table1}.   The shape of the 
free-streamlines can be 
determined by using the {\it hodograph method} \cite{milnethomson,birkhoff}.   
The idea is that while the 
field lines in the $\zeta$-plane may be complicated, the representation of 
the field configuration 
in the $w$ and $H$ planes is quite simple;  a suitable conformal transformation
which maps the $w$ plane onto the $H$ plane would then lead to a relation 
between $w$ and 
$dw/d\zeta$, the solution of which will determine the 
shape of the interface.  Consider first the $w$-plane.  
Since the magnetic field is tangent to both the nose segment and to the 
normal-superconducting interface, the magnitude of the vector potential 
is constant on these segments, as well as on the centerline shown dashed
in Fig.~\ref{diagram}.  Far from the 
sample (at a fixed $z\rightarrow \infty$), we know that 
$A_{y}\sim H_a x$; therefore,
for laminae separated by a distance $a$, the vector potential on the 
interface of the $n-{\rm th}$ lamina is $A_{y} = H_a na $. The 
potential $\phi$ behaves as $-zH_a$ as $z\rightarrow \infty$, hence the
lamina correspond to the positive half of the $\phi$ axis.   
The field configuration in the $w$-plane is shown in Fig.~\ref{wsigmaplane}(a). 
Next we consider the $H$-plane;  this is simpler if we introduce the 
normalized magnetic field 
\begin{equation}
\eta =  {H\over H_n}  = - {1\over H_n} {dw \over d\zeta}.
\label{eta}
\end{equation}  

\dofloatfig
\fi

\begin{figure}
\epsfxsize=3.0 truein
\centerline{\epsffile{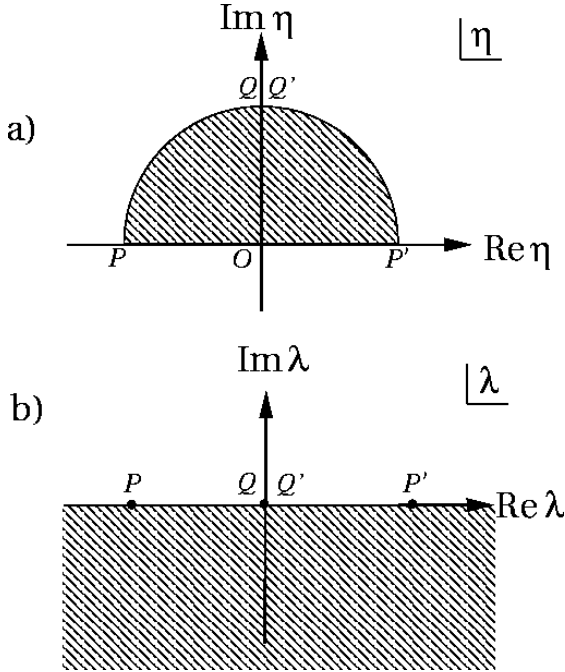}}
\smallskip
\caption[]{The $\eta$- and $\lambda$-planes.}
\label{etalambdaplane}
\end{figure}

On the interface we have $|\eta|=1$, and we define $\theta$ to be the
tangent angle (as in Fig.~\ref{diagram}) as $\eta=-e^{-i\theta}$ and
only the phase $\theta$ changes 
as we traverse the interface from the point $P_n$ to $Q_n$; 
this produces a semi-circle
in the $\eta$-plane.  On the nose $\eta_x = 0$, so that the segment $P_n'P_n$ 
maps onto a horizontal straight line in the $\eta$ plane.  Therefore, 
the region 
{\it exterior} to the superconducting laminae maps onto the {\it interior} of a 
semi-circle, as shown in Fig.~\ref{etalambdaplane}(a)
(the {\it hodograph} for the field).  

We now need to map the $w$-plane onto the $\eta$-plane.  This is most 
easily carried out by the following sequence of transformations. 
First, we take care of the periodicity in the $w$-plane by mapping 
it onto the $\sigma$-plane with the transformation 
\begin{equation}
\sigma = e^{2\pi w/H_a a} - 1,
\label{sigma}
\end{equation}
so that the $\sigma$-plane has a single cut along the positive real 
axis, as shown in Fig.~\ref{wsigmaplane}(b).  Next, map the $\sigma$-plane  
onto the lower half of the $\lambda$-plane using
\begin{equation}
\lambda = \left( {k^2 \over \sigma}\right)^{1/2},
\label{lambda}
\end{equation}
as shown in Fig.~\ref{etalambdaplane}(b).  Here $k$ is a constant to be
determined from the boundary conditions.   The next step is to 
map the $\eta$-plane onto the $\lambda$-plane using the 
Joukowsky transformation
\begin{equation}
\lambda = {1\over 2} \left( {1\over \eta} + \eta \right). 
\label{joukowsky}
\end{equation}
To determine $k$, we notice that as $z\rightarrow \infty$, 
$w\rightarrow -\infty$, so that $\sigma= -1$ and 
$\eta = H_a/H_n$.  Introducing the notation $h_a \equiv H_a/H_n$, we then 
have 
\begin{equation}
k^2 = {(1-h_a^2)^2 \over 4 h_a^2}.
\label{k}
\end{equation}

\dofloatfig
\fi

Combining Eqs.~(\ref{sigma})-(\ref{k}), we have for the 
complex potential 
\begin{equation}
w = {H_a a \over 2\pi} \ln\left[ {(\eta^2 + h_a^2)(\eta^2 + h_a^{-2}) 
                              \over (1+\eta^2)^2} \right].
\label{w_final}
\end{equation}
To determine the shape of the lamina, we use Eq.~(\ref{eta}) to 
write 
\begin{eqnarray}
{d\zeta\over d\eta} &= &  {d\zeta \over d w} {d w \over d \eta} =  
-{1\over H_n}{1\over\eta} {d w(\eta) \over d\eta} \nonumber \\
      & = &  - {a\over \pi} { (1-h_a^2)^2 \over h_a} 
       { 1-\eta^2  \over (\eta^2 + h_a^{-2})(\eta^2 + h_a^2)(\eta^2+1)}.
\label{dzeta}
\end{eqnarray}
The integral can be performed by using a partial fraction expansion. 
The constant of integration is determined by recognizing that 
when $\eta = 0$, $z=0$.  We finally have 
\begin{eqnarray}
\zeta(\eta) &=& -{ia h_a \over \pi} \Biggl[ 
{1\over 2h_a} \ln\left( {h_a- i\eta \over h_a + i\eta}\right) 
 + {h_a\over 2} \ln\left( {h_a^{-1}- i\eta \over h_a^{-1} + i\eta}\right)
\nonumber \\
&&\qquad\qquad -  \ln\left( {1- i\eta \over 1  + i\eta}\right) \Biggr]. 
\label{final}
\end{eqnarray}
This equation implicitly determines the magnetic field $\eta$ as a 
function of position $\zeta$.  The particular parameterization is 
different from Landau's \cite{Landau,LandauLifshitz}, but can be shown to 
be equivalent.   This result also appears to coincide with result of 
Fortini and Paumier \cite{fortini}, although the method of derivation 
is entirely different.  The equivalent fluid problem would be the separated 
flow past a plate placed symmetrically in a channel; the solution to this 
problem is given in Ref.~\cite{birkhoff}, p. 39, Eq. (25a), which is the 
same as Eq.~(\ref{final}) above.  

To calculate the lamina shape, recall that on the normal-superconducting
interface $\eta = -e^{-i\theta}$, with $\theta$ the tangent angle on the 
interface.  Substituting into Eq.~(\ref{final}), and separating real and 
imaginary parts, after a great deal of algebra we obtain the following 
two parametric equations for the position of the interface:
\begin{eqnarray}
x(\theta) &=& {a \over 2} \Biggl[1-h_a - {\left(1-h_a^2\right)\over \pi} 
\tan^{-1} \left( {2 h_a \cos\theta \over 1-h_a^2}  \right)\Biggr]~,\nonumber \\
z(\theta) &=& {a \over 4\pi} 
\Biggl[\left(1+h_a^2\right)
 \ln\left({1+h_a^2 + 2 h_a \sin\theta \over 1+h_a^2 - 2 h_a \sin\theta }
    \right)\nonumber \\
&&\qquad\qquad - 4h_a \ln\left( {\cos\theta \over 1 - \sin\theta}\right) \Biggr].
\label{z}
\end{eqnarray}

\begin{figure}
\epsfxsize=3.3 truein
\centerline{\epsffile{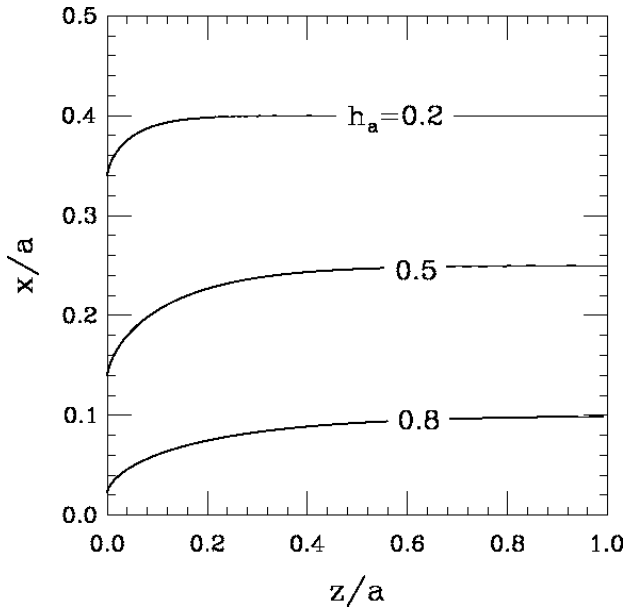}}
\smallskip
\caption[]{Lamina shapes calculated from Eq. (\ref{z}) for several
values of the reduced field $h_a$.}
\label{shapes}
\end{figure}

Using these parametric equations we can calculate the lamina shapes for 
different applied fields $h_a$; some representative results are given in 
Fig.~\ref{shapes}.  The width $2b$ of the nose can now be determined from 
(\ref{z}) by setting $\theta = 0$, with the result
\begin{equation}
b = {a\over 2} (1- h_a) - {a \over 2\pi}(1-h_a^2) 
               \sin^{-1}\left( {2 h_a\over 1+h_a^2}\right) . 
\label{b}
\end{equation}

\dofloatfig
\fi

\subsubsection{Energy of the laminar structure} 

The total free energy of the laminar state has contributions from the 
condensation energy, the magnetic field, and the SN
interfaces. The condensation energy is just 
$-H_{c}^2/8\pi$ times the volume occupied by the superconducting phase. 
If the normal-superconducting interfaces did not bend at all, this energy 
would be $-(H_c^2/8\pi) (N a_s) L_y d$, with $N$ the total number of laminae, 
$L_y$ the length of the sample in the $y$-direction,  and $d$ the 
thickness of the sample  [note that the sample has a total area 
$A= (Na) L_y$]. To this we must add the condensation energy lost 
due to the thinning of the superconducting regions near the sample surfaces. 
The condensation energy per unit area is then 
\begin{equation}
{E_{\rm c} \over A} = - {H_c^2 d\over 8\pi} {a_s \over a}   
      + {4\over a} {H_c^2  \over 8\pi} \int_{-\infty}^0 
             \left[ {a_s\over 2} - x(z)\right] dz, 
\label{condense1}
\end{equation}
where the factor of $4$ in front of the integral accounts for the $4$ corners 
of the lamina.  The integral may be written as an integral over the 
tangent angle $\theta$:
\begin{eqnarray}
&&\int_{-\infty}^0 \left[ {a_s\over 2} - x(z)\right] dz =
  - \int_{0}^{\pi/2}  \left[ {a_s\over 2} - x(\theta)\right] {dz \over d\theta} 
            d\theta \nonumber \\
 &=& {a^2h_a(1-h_a^2)\over 2\pi^2}\! \int_{0}^{\pi/2}\!\!\!\! {d\theta}
    {\tan^{-1}(p\cos\theta)(1-\cos^2\theta) \over \cos\theta (1+p^2\cos^2\theta)}, 
\label{condense2}
\end{eqnarray} 
where the second line was obtained by substituting from (\ref{z}), 
and with $p = 2h_a/(1-h_a^2)$.  After a partial fraction expansion and 
several integrations by parts the integral in (\ref{condense2}) is found to be
\begin{equation}
{\pi \over 2} \left( \ln(p + q) - {q \over p}
\ln q \right)~,
\label{int1}
\end{equation}
with $q=(1+p^2)^{1/2}$.
We then have for the condensation energy 
\begin{equation}
{E_{\rm c} \over A} = - {H_c^2  d \over 8\pi} 
 +  {H_c^2  d \over 8\pi}\left[ {a_n \over a} + {a\over d} f_c(h_a)\right],
\label{condense3}
\end{equation}
with
\begin{eqnarray}
f_c(h) &=& {1-h^2 \over 2\pi} \Bigl[(1+h)^2 \ln (1+h) + (1-h)^2\ln(1-h) \nonumber \\
&&\qquad\qquad - (1+h^2) \ln(1+h^2) \Bigr].
\label{fc}
\end{eqnarray}

\begin{figure}
\epsfxsize=2.8 truein
\centerline{\epsffile{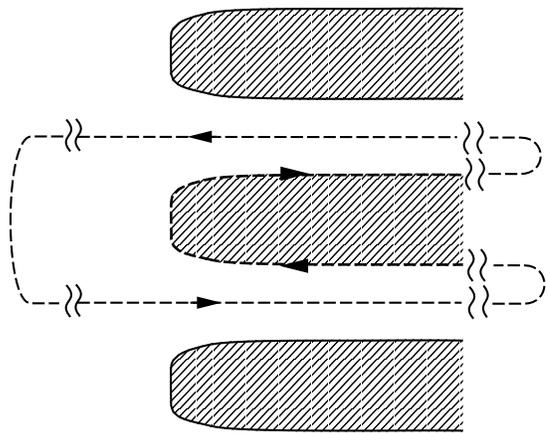}}
\smallskip
\caption[]{Contour for energy calculation.}
\label{contour}
\end{figure}

Next, we need to find the magnetic field energy, obtained by 
integrating $B^2/8\pi$ over all space, both inside and outside the 
sample.   For our periodic structure the field energy per unit area becomes
\begin{equation}
{E_{\rm m} \over A} = {2\over a} \int_{\cal C} {B^2 \over 8\pi}\, dx\, dz,
\label{field1}
\end{equation}
where the factor of 2 accounts for the top and bottom surfaces of the 
sample, and the integral is taken over the area ${\cal C}$ of one cell, as 
shown in Fig.~\ref{contour}.  We can write $B^2 = (\nabla A_y)^2$ in this two 
dimensional geometry, and then use the fact that $\nabla^2 A_y = 0$ to 
write Eq.~(\ref{field1}) as a line integral around the boundary 
$\partial {\cal C}$ of the unit cell:  
\begin{equation}
{E_{\rm m} \over A} = {2\over 8\pi a} \int_{\partial {\cal C}} 
                    A_{y}(s) B_s(s)\, ds,
\label{field2}
\end{equation}
with $A_{y}(s)$ the vector potential on the boundary, and $B_s(s)$ the 
component of the magnetic field which is tangent to the boundary.  

\dofloatfig
\fi

The advantage of this representation of the field energy is that 
the vector potential is {\it constant} on the boundaries, and it can therefore
be brought outside the integral.  Let's consider
the various contributions to the integral.  On the 
normal-superconducting interface (the segments $Q'P'$ and $PQ$) 
the integral $\int B_s ds$ vanishes, since the field points in the same 
direction on the left half of the superconducting lamina as on the right half.  
On the midline between the laminae labeled $n-1$ and $n$,
the vector potential is $H_a a (n-1)/2$ while on the next midline up
it is $H_a (n+1)/2$; the integral $\int B_s ds$ is 
equal in magnitude but opposite in sign for these two segments, since the 
integration paths are in opposite directions.  Adding these two 
contributions, and using $B_s = - \partial \phi /\partial s$ we have 
\begin{eqnarray}
{E_{\rm m} \over A} & = & {H_a \over 4\pi} \int_{\partial{\cal C}} 
B_s\, ds \nonumber \\
  & = & {H_a \over 4\pi}\left[\phi(a/2,-d/2)
-\phi(a/2,L_z/2)\right]~,
\label{field3}
\end{eqnarray}
where $L_z$ is some large distance away from the top surface of the 
sample.  The entire calculation of the field energy then reduces to finding
the asymptotic behavior of the scalar potential along one of the 
streamlines (the midline, in this case).  By examining the behavior of
Eq.~(\ref{w_final}) for the complex potential $w$, and Eq.~(\ref{final}) for
the position $\zeta$,  as $\eta\rightarrow 1$ and $\eta\rightarrow h_a$, 
we find the asymptotic behavior $\phi(a/2,L_z/2)\sim - H_a L_z/2 + \phi_+$,
where
\begin{eqnarray}
\phi_+&=& -{H_a a \over 2\pi} \Bigl[\ln 4 
+(1+h_a^2)\ln(1+h_a^2) \nonumber \\
&&\qquad\qquad  - (1+h_a)^2 \ln(1+h_a) \nonumber \\
&&\qquad\qquad - (1-h_a)^2\ln(1-h_a)\Bigr],
\label{phi1}
\end{eqnarray}
and $\phi(a/2,-d/2)\sim - H_n a/2 + \phi_-$, where
\begin{eqnarray}
\phi_-&=& -{H_n a \over 2\pi} \Bigl[(1-h_a)^2\ln(1-h_a) \nonumber \\
&&\qquad\qquad- (1+h_a)^2 \ln(1+h_a)\nonumber \\
&&\qquad\qquad + 2 h_a \ln 4h_a\Bigr]. 
\label{phi2}
\end{eqnarray}
Substituting into (\ref{field3}), we have 
\begin{equation}
{E_{\rm m} \over A}= {H_a^2\over 8\pi} L_z + {H_n H_a\over 8\pi} d 
          + {H_n^2 \over 4\pi} a f_{\rm mag}(h_a),
\label{field4}
\end{equation}
with 
\begin{eqnarray}
f_{\rm mag}(h_a) &=& {h_a \over 2\pi} \Bigl[(1+h_a)^3\ln(1+h_a)\nonumber \\
&&\qquad - (1-h_a)^3\ln(1-h_a) \nonumber \\
&&\qquad - h_a(1+h_a^2)\ln(1+h_a^2) - 2h_a\ln 8h_a \Bigr]~.\nonumber \\
\label{field5}
\end{eqnarray}
The first term in Eq.~(\ref{field4}) is the energy of the external field in 
the absence of the sample, which is of no interest and will be dropped 
from now on.  The second term is the bulk magnetic field energy  for 
a uniformly magnetized sample \cite{magenergy}, and the third term 
is the energy arising from demagnetizing fields (due to the partitioning 
of the sample into domains).  

Finally, we need to calculate the surface energy due to the 
normal-superconducting interfaces.  If $\sigma_{SN}$ is the surface 
tension for the normal-superconducting interface, then the energy for a
single interface is $\sigma_{SN} d L_y$ \cite{tension}.  Since there 
are two interfaces 
per lamina, and $N$ lamina in the sample, the total energy due to the 
interfaces is $2\sigma_{SN} d L_y N = 2\sigma_{SN} A (d/a)$.  
We can introduce a length $\Delta$, which is essentially the width 
of the interfaces, through $\sigma_{SN} = (H_c^2/8\pi)\Delta$; 
then the energy per unit area due to the interfaces is 
\begin{equation}
{E_{\rm int} \over A} = {H_{c}^2\over 8\pi} { 2\Delta d \over a}.
\label{interface}
\end{equation}

\begin{figure}
\epsfxsize=3.3 truein
\centerline{\epsffile{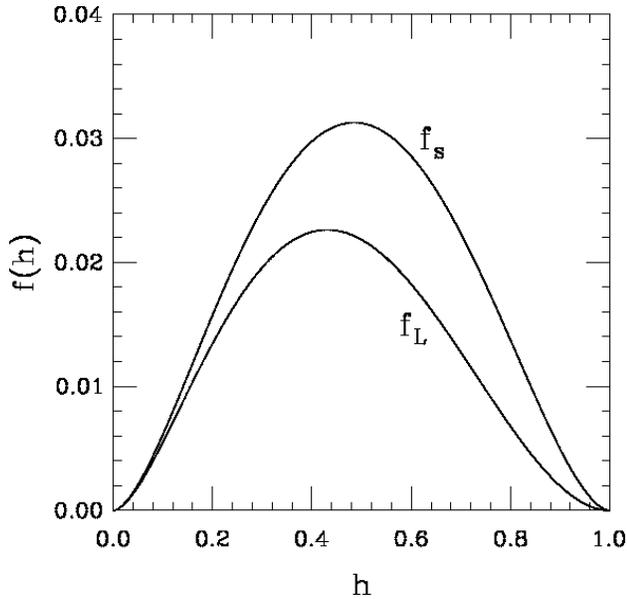}}
\smallskip
\caption[]{The functions $f_{\rm L}(h)$ in Landau's model and $f_s(h)$
in the straight-lamina approximation.}
\label{fh}
\end{figure}

Adding together all of the contributions to the energy, Eqs.~(\ref{condense3}),
(\ref{field4}), and (\ref{interface}), and using the flux conservation 
constraint $H_n a_n = H_a a$, we find 
\begin{eqnarray}
{E\over A} &=& {H_c^2 d \over 8\pi} \Biggl\{-1
+ \left[ {a_n\over a}+ h^2 {a\over a_n}\right]
\nonumber \\
&&\! + 2 \left( {\Delta \over a} + 
  {a\over d} \left[ 2 f_{\rm c}(h_a) + {H_n^2\over H_c^2}
 f_{\rm mag}(h_a) \right] \right)\Biggr\},
\label{total}
\end{eqnarray}
with $h\equiv H_a/H_c$. 
The energy must be minimized with respect to both $a$ and $a_n$.  This 
results in very cumbersome expressions.  Instead, we will minimize the 
first term in brackets with respect to $a_n$, which yields 
$a_n = h a$, so that $H_n=H_c$ (and $h_a = h$).  This is the result used by 
Landau, which is reasonably accurate as long as the surface 
and demagnetizing energies are small.  If we substitute this back 
into the energy, we obtain
\begin{equation}
{E\over A} = - {H_c^2 d \over 8\pi} + {H_c H_a d \over 4\pi} 
 + {H_c^2 d \over 4\pi} \left[ {\Delta \over a} +  {a\over d} f_{\rm L}(h)\right],
\label{total2}
\end{equation}
with 
\begin{eqnarray}
f_{\rm L}(h)&=&2f_{\rm c}(h) + f_{\rm mag}(h) \nonumber \\
&=&{1\over 4\pi}\Bigl[(1+h)^4\ln(1+h) + (1-h)^4\ln(1-h)\nonumber \\ 
&&\qquad - (1+h^2)^2\ln(1+h^2) - 4h^2\ln 8h \Bigr].
\label{f}
\end{eqnarray}
This function is plotted in Fig.~\ref{fh}. 
Its asymptotic behavior as $h\to 0$ is of interest in comparison with
other approaches discussed below, and has the form
\begin{equation}
f_{\rm L}(h)\simeq {h^2\over \pi}\ln\left({0.56\over h}\right)~.
\label{f_asym}
\end{equation}

Finally, the equilibrium laminar period is obtained
simply by minimizing  with respect to $a$, yielding
\begin{equation}
a^* = \left[ {\Delta d \over f(h) }\right]^{1/2}.
\label{a}
\end{equation}

\dofloatfig
\fi

\noindent This implies that the spacing diverges for small $h$ as
\begin{equation}
a^*\simeq {(\Delta d)^{1/2}\over h}
\left[{\pi\over \ln\left(0.56/h\right)}\right]^{1/2}~.
\label{a_smallh}
\end{equation}

\subsection{Energy of a straight-lamina model} 

\begin{figure}
\epsfxsize=3.3 truein
\centerline{\epsffile{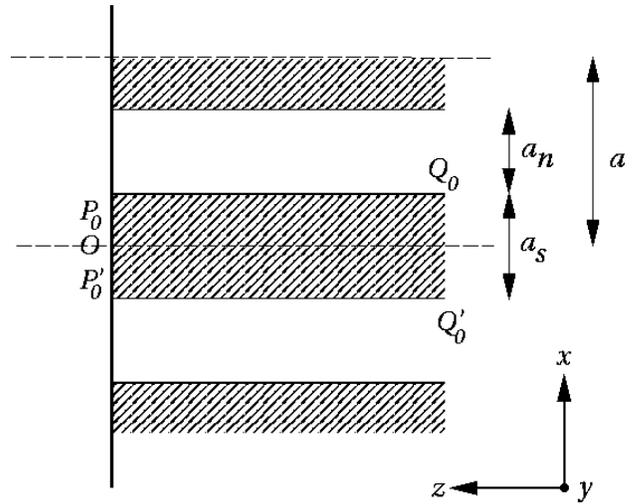}}
\smallskip
\caption[]{Geometry of the straight-laminae model.}
\label{straight}
\end{figure}

It is useful to compare the results from the Landau model to an 
alternative {\it straight-lamina model},  which is illustrated in 
Fig.~\ref{straight}.  In this model the field is still tangent to the 
normal-superconducting interface, but the magnitude of the field is 
not constant along the interface.  The magnetic field and the complex potential 
can be obtained using standard conformal mapping methods; this problem is 
equivalent that of an ideal fluid flowing in a channel with an abrupt 
step \cite{milnethomson}.  The solution is 
\begin{equation}
w= {H_a a \over 2\pi} \ln\left( {\eta^2 +1 \over \eta^2 + h_a^2}\right),
\label{straight1}
\end{equation}
\begin{equation}
\zeta= {a_s \over 2} + {i a h_a \over 2\pi} 
   \left[ \ln\left( {\eta+i  \over \eta-i}\right) - {1\over h_a} \ln\left(
     {\eta + ih_a \over \eta - ih_a}\right) \right].
\label{straight2}
\end{equation}

Using this solution it is possible to calculate the total energy, as in the 
Landau model.  Making the simplifying assumption that $H_n = H_c$, we 
obtain 
\begin{equation}
{E\over A} = - {H_c^2 d \over 8\pi} + {H_c H_a d\over 4\pi} 
   + {H_c^2 d \over 4\pi} \left[ {\Delta \over a} + {a\over d} f_s(h)\right],
\label{straight3}
\end{equation}
 
\dofloatfig
\fi

\noindent with
\begin{eqnarray}
f_s(h) &=& {h\over 2\pi}\Bigl[(1+h)^2 \ln(1+h) - (1-h)^2\ln(1-h)\nonumber \\
&&\qquad\qquad - 
          2h\ln 4h\Bigr]. 
\label{straight4}
\end{eqnarray}
This function is plotted in Fig.~\ref{fh} for comparison with Landau's result
(\ref{f}).  At small $h$ it behaves as
\begin{equation}
f_s(h)\simeq {h^2\over \pi}\ln\left({0.68\over h}\right)~,
\label{fs_asym}
\end{equation}
which is very close to the Landau result (\ref{f_asym}).
We see that the total energy 
for this model has the same qualitative dependence upon the lamina spacing 
$a$ as the Landau model, although the Landau model has a lower energy for 
any reduced field $h$.

\subsection{Laminae shapes in finite-thickness plates}

In the analyses above we have assumed that the slab of superconductor
is sufficiently thick that the shapes of the lamina walls can be computed
as for a semi-infinite slab.  When the thickness $d$ becomes small enough
the wall shapes will change.  From the asymptotic behavior of the
semi-infinite solutions (\ref{z}), we deduce that the thickness approaches
its asymptotic values for $z\to -\infty$ as
\begin{equation}
{a\over 2}(1-h_a)-x(z) \sim {4h_a\over \pi} \exp(\pi z/ah_a)~.
\end{equation}
The decay length $a h_a/\pi$ should then determine when finite-slab
thickness effects become important.  The asymptotic results
(\ref{f_asym}) and (\ref{fs_asym}) show that the product $ah$ vanishes
very slowly (logarithmically) as $h\to 0$, so that while such
finite-thickness effects become important in that limit,
practically the relevant fields are very small.  From the
asymptotics, the crossover field $h_x$ for $a h_a/\pi=d$ is on the order of
\begin{equation}
h_x\sim 0.56 \exp(-\Delta/\pi d)
\label{h_cross}
\end{equation}
For $h\le h_x$ the slab thickness has no significant effect on the
domain wall shapes.
In the fluid dynamical analogy, the finite-thickness slab calculation
is equivalent to so-called Riabouchinsky flows around {\it two} plates,
the details of which the interested reader will find in
standard references \cite{birkhoff}.  

\section{The current loop model}

The analysis of the laminar state in the previous section has shown that
in accounting for the flaring of the normal domains Landau's free-boundary
approach yields a lower energy structure than a model with straight
walls.  But the analytical and numerical differences between the
two approaches are relatively minor.  In both models the supercurrents
flow along the SN interfaces {\it and} on the top and bottom surfaces
of the sample \cite{Narayan,Yu}.  Just as the magnetic field 
in a solenoid is more nearly
uniform when it is a tall thin cylinder than when it is short and wide,
so too do we expect that the contributions from circulating currents
along the SN interfaces will dominate when the flux domains are
narrow and tall, at low applied fields $h$.  This suggests that
the basic physics of the laminar state can be understood by
considering those circulating currents alone.  We develop this
current-loop model in the present section and show that it 
rather accurately reproduces the results of Landau's model.
This calibration is an important test of an approach that
can easily be generalized to SN interfaces of arbitrary shape.

\subsection{Energetics of the current loop model}

For the purposes of this model, the intermediate state is
described in macroscopic terms.  As above, it has
thickness $d$, total area $A$, and volume $V=Ad$, but now the
SN interfaces encircling each of the normal regions belong to a set
$\left\{{\cal D}_i\right\}$ each with
area $A_i$ and perimeter $L_i$.  
The two phases occupy volumes $V_s$ and $V_n=d\sum_iA_i$, with $V_s+V_n=V$.  
Parameterizing each boundary by arclength,
the position vectors of the interfaces 
are ${\bf r}_{i}(s)$.  As in the straight-lamina model,
we assume that ${\bf r}_{i}$ is
independent of $z$, neglecting
the flaring of the domain walls near the film surfaces that is
seen in Fig. \ref{shapes}.

The total energy of the system is as before a sum of the
condensation energy $E_c$, the interfacial energy $E_{\rm int}$,
and the magnetic field energy $E_{\rm m}$,
\begin{equation}
E[\{ {\bf r}_i\}] = E_c + E_{\rm int} + E_{\rm m}.
\label{totalCL}
\end{equation}
\dofloatfig
\begin{figure}
\epsfxsize=3.3 truein
\centerline{\epsffile{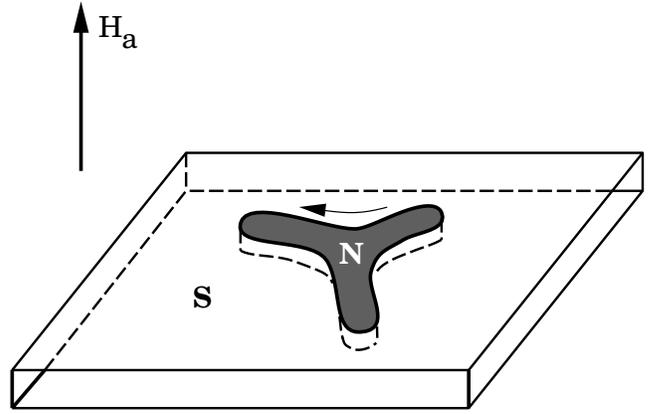}}
\smallskip
\caption[]{A current loop.}
\label{loop_fig}
\end{figure} 	
\fi
\noindent And as before,
\begin{equation}
E_{\rm c}=V{H_c^2\over 8\pi}\rho_n ~,
\ \ \ E_{\rm int}= {H_c^2\over 8\pi}\Delta d \sum_i L_i~,
\label{bulk_energy}
\end{equation}
where $\rho_n =  A_n/A$ is the area fraction for the normal 
phase, and where $E_{\rm c}$ is measured with respect to the purely
superconducting state.
Global flux conservation relates the field in the normal regions
to the applied field, $H_n=H_a/\rho_n$,
and by the tangential continuity of ${\bf H}$ across
a S-N interface the field in the superconducting region
is $H_s =H_n $.  The 
superconducting regions are perfectly diamagnetic; they each have
a magnetization ${\bf M} = -(H_{n}/4\pi) {\bf \hat e}_z$,
related in the usual way to the (Meissner) currents that flow
along the SN (and top and bottom sample) boundaries.

We compute the magnetic field energy as a sum of two
contributions, the first of which is that of
the domain magnetizations ${\bf M}$ 
in the presence of the external field ($-\int d^3r{\bf H}_a\cdot {\bf M}$).
The second contribution is the self- and mutual-induction of the current
ribbons.
Expressing these in terms of the macroscopic quantities $\rho_n$, etc.,
and the current-ribbon positions we have
\begin{eqnarray}
E_{\rm m} &=& V{H_aH_n\over 4\pi}\left(1-\rho_n\right)\nonumber \\
&&\!\!-{1\over 2}M^2\!\sum_{i,j}
\int_0^d\!\!\! dz\! \!\int_0^d \!\!\!dz'\!\oint\!ds\oint\!ds'
{{\bf \hat t}_i\cdot{\bf \hat t}_j\over R_{ij}}~,
\label{totalenergy}
\end{eqnarray}
where $M=-H_n/4\pi$.
Here, the vectors ${\bf \hat t}_i={\bf \hat t}_i(s)$ are unit tangents to the
current ribbons and label the direction of the current flow.
By the usual screening processes in superconductors, the direction of
the flow is so as to cancel the applied field in the superconducting
regions and augment it in the normal regions (see Fig. \ref{loop_fig}).  
The scalar product
of the tangent vectors is however invariant under the overall reversal of the
parameterizations ($s\to -s$). 
The current-current interaction is Coulombic, with 
$R_{ij}=\{[{\bf r}_i(s)-{\bf r}_j(s')]^2+(z-z')^2\}^{1/2}$.  
While the $z$ and $z'$ integrals are readily performed (see below),
the more elementary form (\ref{totalenergy}) serves to remind us
that the current-current interactions are in their free-space form.
By performing the $z$ and $z'$ integrals, the magnetic field energy becomes
\begin{eqnarray}
E_{\rm m} &=& V{H_aH_n\over 4\pi}\left(1-\rho_n\right)\nonumber \\
&&\!\!-M^2 d\!\sum_{i,j}
\oint\!ds\oint\!ds'
{\bf \hat t}_i\cdot{\bf \hat t}_j \Phi(R_{ij}/d), 
\label{totalenergy2}
\end{eqnarray}
where now the elementary free boundaries are {\it contours in
the plane}, interacting with the potential 
\begin{eqnarray}
\Phi(R/d) & = & {1\over 2 d} \int_{0}^{d}\! dz \int_{0}^{d} \! dz'
\left[R^2 + \left(z-z'\right)^2\right]^{-1/2} \nonumber \\
 & = & \sinh^{-1}(d/R) + R/d  - \sqrt{1+(R/d)^2}~,
\label{Phi}
\end{eqnarray}
where $R=\vert{\bf R}\vert$ with
${\bf R}(s,s')={\bf r}(s)-{\bf r}(s')$ the in-plane vector between
points labeled by $s$ and $s'$.
As discussed elsewhere \cite{pra}
this potential is Coulombic for $R\gg d$, $\Phi\approx d/(2R)$,
but for $R \ll d$,  $\Phi\approx \ln (2 e^{-1}d/R)$, with the
film thickness $d$ acting as a cutoff.  Note the interesting parallel
with Pearl's interaction potential (\ref{pearl_interaction}) 
among vortices in thin films.
The fact that this interaction potential is in some cases identical and
in others nearly identical to that found in the free-boundary approach
to a number of other systems offers an explanation for the similarity
in their behavior.
Table II summarizes these analogies between the different systems.

\dofloatfig
\begin{figure}
\epsfxsize=3.3 truein
\centerline{\epsffile{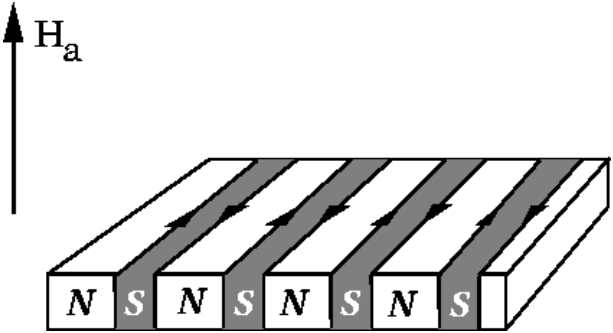}}
\bigskip
\caption[]{The laminar state as a collection of current loops. Arrows
indicate directions of the current, tangential to the SN interfaces.}
\label{laminar_cl}
\end{figure} 	
\fi

\subsection{Current-loop description of the laminar state}

Now we calculate the energy of the laminar state in the CL model.
As before, the periodicity length in the plane is $a$, and
the width of the normal lamina is $a_n$; we will assume that the
pattern is periodic in the $x$-direction. 
The nonlocal magnetic contribution is the only difficult one, and
it proves more convenient to return to the original self-induction
form of the tangent vector coupling, before averaging over the slab
thickness.  It is then easy to introduce a Fourier representation,
which for a uniform laminar structure yields the magnetic field energy 
$E_{\rm m}$ per unit area
\begin{eqnarray}
{E_{\rm m}\over A}&=& -{M^2\over a}\sum_{n=-\infty}^{\infty}
\int_{-\infty}^{\infty}\! dy \int_0^d\! dz\int_0^d\! dz'\nonumber \\
&&\!\!\!\times\!\! \int\!\!{d^3q\over (2\pi)^3}{4\pi\over q^2}
e^{i[nq_xa+q_yy+q_z(z-z')]}\left(1-e^{iq_x a_n}\right)\!.
\!\label{elam2}
\end{eqnarray}
Several straightforward integrations reduce this to 
\begin{eqnarray}
{E_{\rm m}\over A}&=&-{aM^2\over \pi^2}\sum_{n=1}^{\infty}
{\left[1-\cos\left(2\pi n a_n /a\right)\right]\over n^3}\nonumber \\
&&\qquad\qquad\times \left({2\pi n d\over a} + e^{-2\pi d n/a}-1\right).
\label{elam3}
\end{eqnarray}
Note that the last term contains all of the $d$-dependence
of the sum.  The leading contribution in the limit of large slab thickness
is a ``bulk" contribution expressible simply in terms of the
stripe dimensions 
\begin{equation}
\sum_{n=1}^{\infty}
{\left[1-\cos\left(2\pi n a_n/a\right)\right]\over n^2}=
\pi^2{a_n(a-a_n)\over a^2}~.
\label{elam4}
\end{equation}
It follows that the form of the free energy is exactly like that in
Eqs. (\ref{total}) and (\ref{total2}), but with a new function $f$,
\begin{equation}
f_{\rm CL}(h,d,a) = {1\over 2\pi^3} \sum_{n=1}^{\infty}
{\sin^2(n\pi h) \over n^3} \left[ 1 - e^{-2\pi n d/a}\right].
\label{currentloop3}
\end{equation}
As alluded to in the discussion of the laminar shapes in Landau's
calculation, finite-thickness effects show up when $d$ is comparable to
$a$.

\dofloatfig
\begin{figure}
\epsfxsize=3.3 truein
\centerline{\epsffile{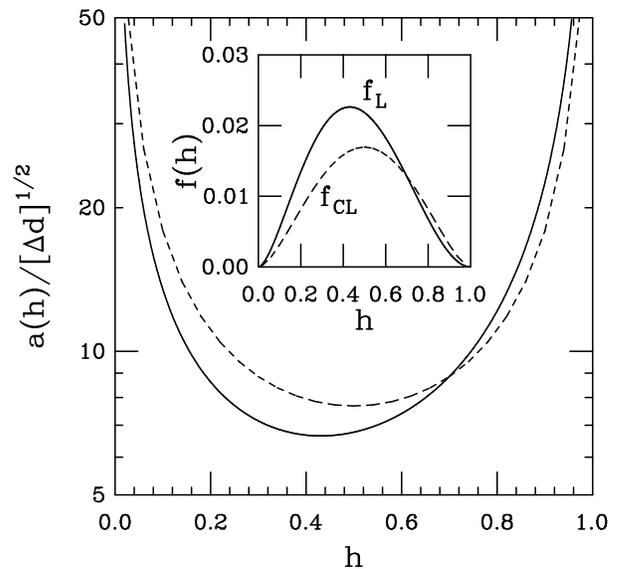}}
\smallskip
\caption[]{Comparison of current loop model and Landau model for
laminar state.   Equilibrium laminar spacing $a(h)$ 
for Landau model (solid) and current-loop model (dashed).
Inset shows the functions $f_L$ (solid) and $f_{CL}$
(dashed).}
\label{fignew}
\end{figure} 	
\fi

There are several noteworthy features of the function
$f_{\rm CL}(h)$, particularly in the limit $d/a \gg 1$ which we consider here.  
First, as shown in Fig. \ref{fignew}, it is rather
close to the Landau function, and hence its implications for the
equilibrium laminar thickness are in reasonable accord with experiment.
Second, analytically it has the same structure as $f_{\rm L}$ as $h\to 0$,
\begin{equation}
f_{\rm CL}(h)\simeq {h^2\over 2\pi}\ln\left({0.71\over h}\right)~.
\label{fCL_limit}
\end{equation}
Third, it has an {\it exact} symmetry under the transformation $h\to 1-h$,
a reflection in part of the straight SN interfaces presumed in the model.
This symmetry is absent in Landau's calculation and in the 
straight-lamina model by virtue of the currents on the
slab faces.  Finally, the form of the magnetic field energy in 
Eq.~(\ref{elam3}) is identical to the field energy of 
a stripe array in a thin ferromagnetic film \cite{kooy60,cape71}.

\subsection{A dynamical model}

The energetic competition between surface tension and self-induction present
in the current-loop model appears in a number of other
contexts (Table II), where it has been 
shown to produce also a rich dynamical behavior.  
While the precise connection between the Young-Laplace and Biot-Savart
forces and the interface dynamics depends on the physical setting 
(e.g. Hele-Shaw flow with Darcy's Law, surfactant monolayers at
the air-water interface with coupling to the fluid phase, 
reaction-diffusion systems), the
phenomenon of repeated branching instabilities producing disordered
lamellar structures is ubiquitous.  This suggests that much can
be learned by studying the very simplest dynamical law for
interface motion, the {\it local dissipation model} \cite{pra},
in which a local viscous drag acting at
the interface balances the local pressure difference, the
latter computed as a functional derivative as in Eq. (\ref{super_var}).  

As a first step toward a full study of the {\it many-interface}
current-loop model of the 
intermediate state, we study here the simplest mean-field
description of a {\it single} current loop.  That loop is
assigned to a cell (analogous to a Wigner-Seitz cell) 
of area $A_{\rm cell}$ from which we compute the area fraction
$\rho_n=A_n/A_{\rm cell}$.
In this approximate description, the self-induction of the
loop is retained in its full form, but the mutual induction
term in the energy associated with all other loops
only contributes a bulk energy term like that seen in
the laminar calculation (\ref{elam4}).
Moreover, the amplitude of the circulating currents is taken to
be set by $H_c$ rather than the actual local field.  This is
equivalent to assuming that the actual area fraction departs little
from its equilibrium value.
The system is then characterized by a single geometrical parameter
$p\equiv 2R_{\rm equiv}/d$ describing the aspect ratio, where $R_{\rm equiv}$
is the radius of the circle whose area is that of the initial condition,
and a single energetic parameter, the reduced magnetic field $h$.  All other
parameters simply rescale time.

Within this model, the normal component of the interface velocity is found by
functional differentiation:
\begin{eqnarray}
{\bf \hat n} \cdot {\bf r}_t(s) &=&
{H_c^2d\over 8\pi\eta}\Biggl\{\Pi-\Delta{\cal K}(s) \nonumber \\
&&\qquad -{1\over 2\pi d}\oint\!ds'
{\bf \hat R}\times{\bf \hat t}(s') \Psi(R/d)\Biggr\}~,
\label{eom}
\end{eqnarray}
with ${\cal K}(s)$ the curvature, 
and $\Psi(\xi)=\Phi'(\xi)= 1-(1+\xi^{-2})^{1/2}$ is the generalization of the Coulombic 
form of the Biot-Savart force to finite-thickness slabs \cite{pra}.
Finally, the pressure term is
\begin{equation}
\Pi = h^2/\rho_{n}^2-1~.
\label{pressure}
\end{equation}
The kinetic coefficient $\eta$ may be estimated \cite{asymptotics}
from results on the bulk properties of strongly type-I systems,
\begin{equation}
\eta = {H_c^2d \Delta\over 8\pi}{\pi \hbar\over 8 k_BT_c \xi_0^2}~,
\label{eta_estimate}
\end{equation}
where again $T_c$ is the critical temperature and 
$\xi_0$ is the bare correlation length \cite{abrikosovbook}.

A contour dynamics such as (\ref{eom}) is readily generalized to
account for surface tension anisotropy, a material feature that has
long been suggested to play a role in the morphology of the
intermediate state patterns \cite{Huebener}, as it does in
problems such as dendritic growth \cite{dendrite}.  When the anisotropy
is $q$-fold, the parameter $\Delta$ has the form
\begin{equation}
\Delta=\Delta_0\left[1+\epsilon\cos\left(q\theta\right)\right]~.
\label{delta_anisotropy}
\end{equation}
Typical experiments show a $q=4$ or $q=6$ anisotropy \cite{bodmer}.
Our intuition suggests that the variation of $\sigma_{SN}$ through
$\Delta$ will bias instabilities toward $q$-fold symmetry, and
lead to preferred orientations of flux stripes produced
from those instabilities.

\subsection{Instabilities; numerical studies}

Two regular geometries of flux domains have historically
been of interest: circles and stripes.  In the next section we
consider in detail the stability of stripes and stripe arrays (the
laminar state); here we focus on fingering and branching instabilities
of circular domains.  Since linear stability analyses 
for circular interfaces have been presented in detail elsewhere 
in the context of closely related models 
\cite{science,pra,pre,LeeMcConnell,jpc,turingpre},
we will not repeat them here in detail.  Two important qualitative results 
from those studies are that for a given size (domain radius and slab thickness)
(i) there exists a critical applied field below which the circle is
stable and above which azimuthal modes become active, and (ii)
instabilities of increasing mode number occur with ever larger applied field.
These properties may be illustrated through numerical studies of the
contour dynamics, which allow us to see the highly nonlinear regime
far beyond the instabilities.

\dofloatfig
\begin{figure}
\epsfxsize=3.3 truein
\centerline{\epsffile{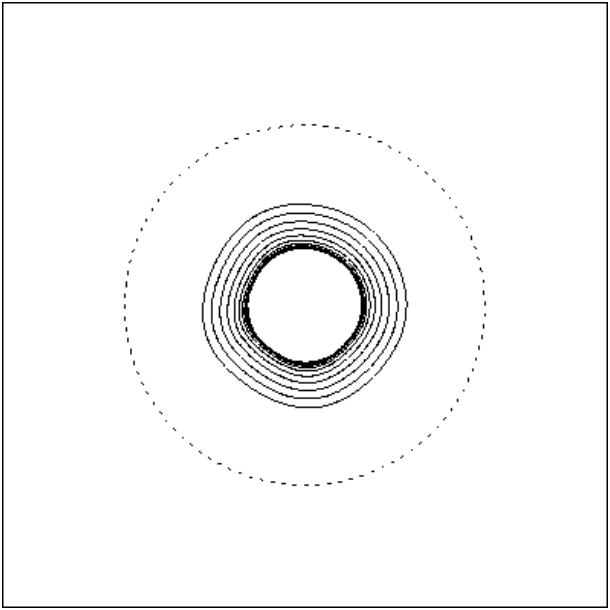}}
\smallskip
\caption[]{Results from numerical studies of the current-loop model.
Relaxation of a circular initial condition to a stable circular final
state of different radius. Dashed line shows the
area of the unit cell. Parameters are $p=5.0$ and $h=0.1$.}
\label{sim0}
\end{figure} 	
\fi

An efficient numerical method for studying this nonlocal interface dynamics
has been described in considerable detail elsewhere \cite{jpc,turingpre}.  
It uses pseudo-spectral techniques to solve for the time evolution of
the local tangent angle $\theta(s)$, from which the ($x(s),y(s)$) coordinates
of the interface are computed by basic differential geometry.   
For the purposes of verifying the analytical stability results as well as
investigating such phenomena as mode competition, the initial contour
is given a curvature ${\cal K}$ perturbed from that of a circle,
\begin{equation}
{\cal K}(\alpha)={1\over R_0}+\sum_{n=2}\left[a_n\cos\left(n\alpha\right)
+b_n\sin\left(n\alpha\right)\right]~,
\label{kappa_perturb}
\end{equation}
where $R_0$ is the unperturbed radius and $\alpha=s/R_0$.

\begin{figure}
\epsfxsize=3.3 truein
\centerline{\epsffile{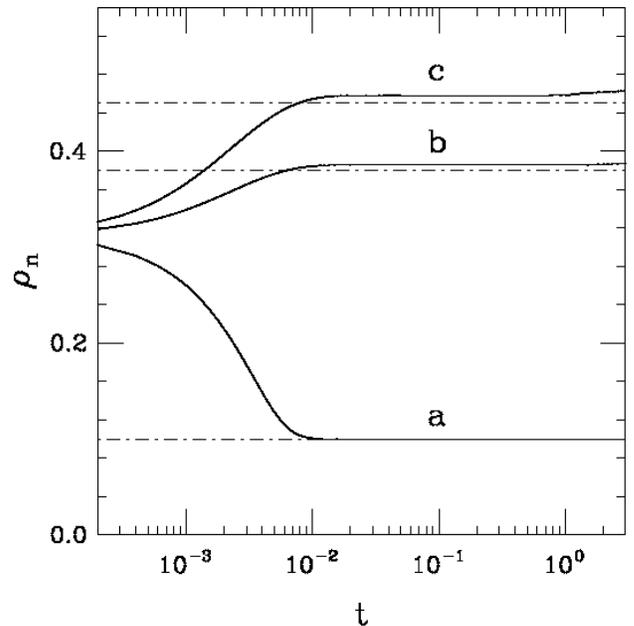}}
\smallskip
\caption[]{Time evolution of the normal area fraction $\rho_n$
for a single domain governed by the contour dynamics (\protect{\ref{eom}}).
Curves (a), (b), and (c) correspond to Figs. \protect{\ref{sim0}},
\protect{\ref{sim2}}, and \protect{\ref{sim3}}.  Dashed lines 
indicate the relation $\rho_n=h$ determined by the bulk energetic
contributions alone.}
\label{sim1}
\end{figure}

There are three basic phenomena that may be illustrated with
the contour dynamics.  The first, shown in Fig. \ref{sim0}, 
is the relaxation of a weakly perturbed circular 
domain whose initial area fraction is not the equilibrium
value.  This {\it stable} relaxation to a circle can occur if the
applied field $h$ is below the instability value at 
the aspect ratio of interest.  The
asymptotic area fraction at long-times is $\rho_n\simeq h$,
apart from a small correction due to surface tension.
The approach of $\rho_n$ to this limiting value is shown in
Fig. \ref{sim1}.

\dofloatfig
\fi

A second phenomenon occurs at higher $h$, and is the elementary 
{\it elongational} instability of a
circular flux domain, as illustrated in Fig. \ref{sim1}.
Again the area fraction evolves toward $\rho_n=h$,
but now the deviation is significant due to larger contributions 
from the Biot-Savart integral.
This shape evolution shows one the means by which finite 
stripes may form in the intermediate state.

\dofloatfig
\begin{figure}
\epsfxsize=3.3 truein
\centerline{\epsffile{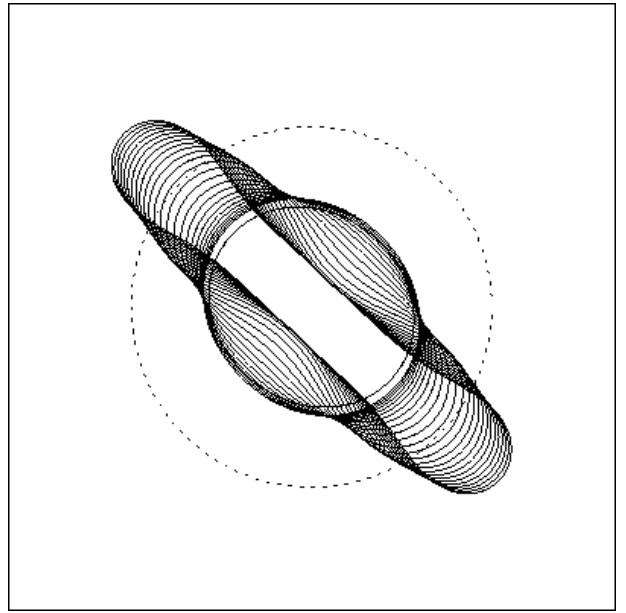}}
\smallskip
\caption[]{Elongational instability of a circular flux domain, with
aspect ratio as in Fig. \ref{sim0}, but $h=0.38$.}
\label{sim2}
\end{figure} 	
\fi

The curious feature of bulbous tips to the stripe
is a rather common observation in dipolar systems.  It is suggestive
that the instability is in some sense related to evolution
toward the fissioning of the original circle into two smaller ones.  
Energetic arguments based on this picture show that it rather 
accurately predicts the onset of this instability \cite{jpc}.

The third phenomenon of interest is the {\it branching} instability that occurs
for still higher values of $h$, as shown in Fig. \ref{sim2}.  The
initial condition for this simulation was a circle perturbed with 
a small amplitude mode of azimuthal number $3$.  Rapid growth of
that mode is followed by relaxation to ``arms" of rather uniform width.
The angles of the three ``arms" forming the vertex are close to
$120^\circ$, as is typical in systems governed by surface tension.

\begin{figure}
\epsfxsize=3.3 truein
\centerline{\epsffile{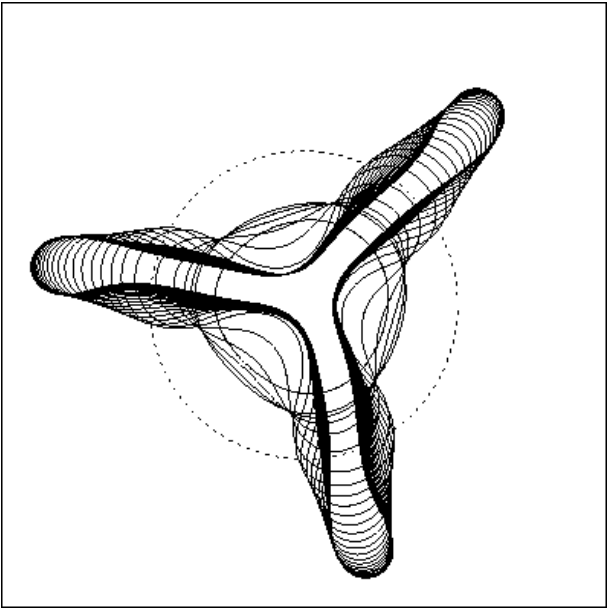}}
\smallskip
\caption[]{Numerical study of branching instability, with $h=0.45$.}
\label{sim3}
\end{figure}

We conclude from these studies that a physical mechanism to produce
the branched and fingered {\it stationary} shapes of flux domains in the 
intermediate state is the mechanical instability illustrated in
Figures \ref{sim2} and \ref{sim3}.

\dofloatfig
\fi

\begin{figure}
\epsfxsize=3.3 truein
\centerline{\epsffile{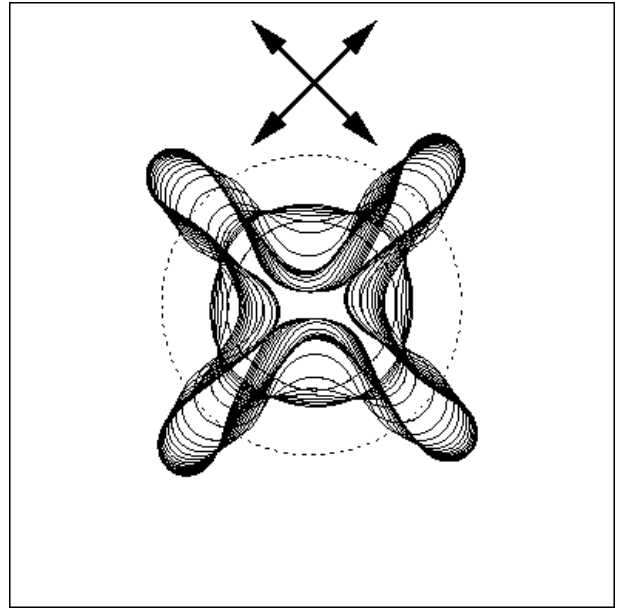}}
\smallskip
\caption[]{Numerical study of branching instability, with $h=0.45$
and surface tension anisotropy ($\epsilon=0.1$, $q=4$).  Arrows indicate
easy axes of low surface tension.}
\label{sim4}
\end{figure}

Finally, Fig. \ref{sim4} shows the effects of 
surface tension anisotropy on the branching instability of
the same initial condition as in Fig. \protect{\ref{sim3}}.
While the time evolution first 
produces a four-fold vertex, it subsequently fissions into
two three-fold vertices that move away from each other.  
The branches of the pattern have oriented themselves with
respect to the low-tension directions determined by
the anisotropy (indicated by arrows).
The instability of vertices of higher order than three is
a common feature of dipolar systems.

\dofloatfig
\fi

\section{Perturbations around the laminar state}

In the absence of any in-plane component to the applied magnetic field,
flux domains in the intermediate state often have the shape of buckled
lamina.  The wavelength is typically larger than the stripe width, as is
the amplitude of the modulation.  The conformal mapping algorithm for
the laminar state is not generalizable to treat such truly three-dimensional
structures, and there does not appear to have been any stability calculation
of the laminar state.  In the following sections we compute both the
stability and elastic properties of flux stripes, making connections with
pattern forming properties in other systems.

\subsection{Energy and stability of a single flux stripe}

\begin{figure}
\epsfxsize=3.3 truein
\centerline{\epsffile{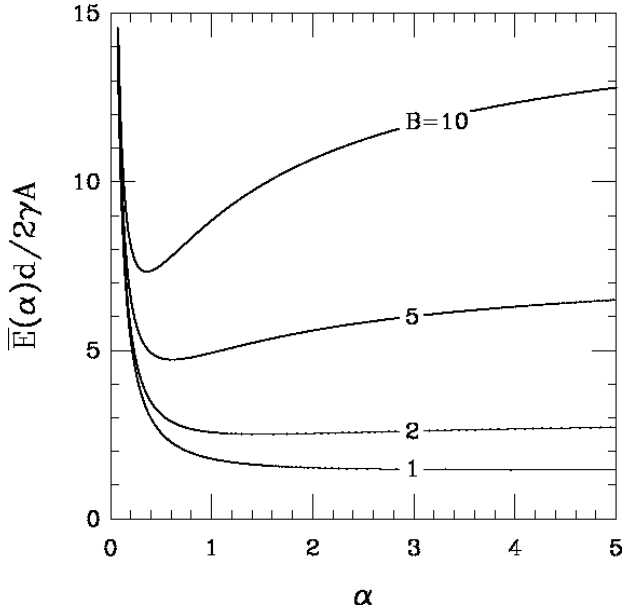}}
\smallskip
\caption[]{Stripe energy density as a function of stripe width, for various magnetic
Bond numbers.}
\label{stripe_fig}
\end{figure}

Consider first the properties of a single flux stripe as described by the
current-loop model. 
If the stripe has width $w$ in the $x$-direction, a length $l$ in the 
$y$-direction, and the plate spacing is $d$,
then by considering the self- and mutual induction of the currents flowing
along the edges the reduced energy $\tilde{E}\equiv  E /2\sigma_{SN} A$
per unit area is 
\begin{equation}
\tilde{E}={1\over \alpha}
-{ N_B \over \alpha}\int_0^{\infty}\! d\xi \left[\Phi(\xi)-
\Phi(\sqrt{\xi^2+\alpha^2})\right]~,
\label{stripe_energy1}
\end{equation}
where $\alpha=w/d$
and $N_B=2M^2d/\sigma_{SN}$ is the dimensionless magnetic {\it Bond} number.
The integrals are standard and yield the result
\begin{eqnarray}
\tilde{E}& =& {1\over \alpha} + { N_B\over 4\alpha}\left[\alpha^2\ln(1+\alpha^{-2})
  \right.\nonumber \\
 & & \ \ \left. +4\alpha\tan^{-1}\alpha -\ln(1+\alpha^2) \right].
\label{stripe_energy2}
\end{eqnarray}
Figure \ref{stripe_fig} shows the stripe energy as a function of
its width for various Bond numbers.  We see that the minimum of
this energy becomes sharper as $N_B$ increases.
Minimizing $\tilde{E}$ with respect to $\alpha$ (hence with respect to
the width) at fixed area $A$ yields a relation between $\alpha$ and $N_B$,
\begin{equation}
1-{N_B\over 4}\left[\alpha^2\ln(1+\alpha^{-2})+\ln(1+\alpha^2)\right]=0~.
\label{Bond_crit}
\end{equation}

\begin{figure}
\epsfxsize=3.3 truein
\centerline{\epsffile{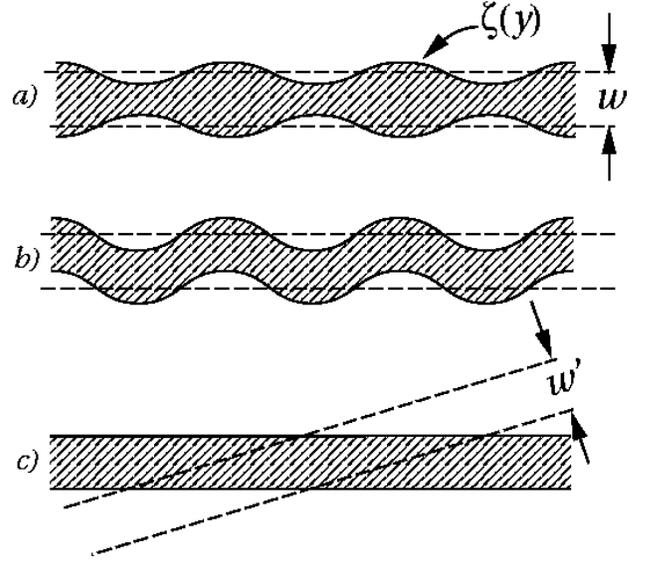}}
\smallskip
\caption[]{Peristaltic (a) and serpentine (b) perturbations of a flux stripe.
(c) Illustration of the change in stripe width upon a uniform rotation.}
\label{stripe_pert_fig}
\end{figure}

Now we connect this result to the stability analysis of the
stripe.  As shown in Fig. \ref{stripe_pert_fig}(a) and (b) there are 
two classes of 
small distortions we must consider.  The first (``peristaltic") involves
antisymmetric perturbations and changes the local stripe width.
This will be of higher energy than the symmetric (or ``serpentine")
distortions of Fig. \ref{stripe_pert_fig}(b) which preserve the width. 
It is most convenient to calculate the linearized {\it force}
acting on the interface, for which we use the result quoted in
Eq. (\ref{eom}), written out explicitly, 
\begin{eqnarray}
-\gamma {\cal K} &+& 2M^2\oint \! ds'
{\left({\bf r}(s)-{\bf r}(s')\right)\over
\vert {\bf r}(s)-{\bf r}(s')\vert}\times \hat {\bf t}(s')\nonumber \\
&&\qquad\qquad\times \left\{\sqrt{1+{d^2\over \vert {\bf r}(s)-{\bf r}(s')\vert^2}}
-1 \right\}~.\label{normal_velocity}
\end{eqnarray}

\dofloatfig
\fi

The details of this perturbation analysis are given in Appendix A.
After considerable algebra we obtain the force associated with
a monochromatic perturbation of reduced wavevector $q=dk$; for the 
serpentine perturbation, 
\begin{eqnarray}
F_s(q)&=&q^2-2N_B\Bigl\{\gamma_E+\ln\left({\alpha q\over 2\sqrt{1+\alpha^2}}\right)
+K_0(q)\nonumber \\
&&\qquad\qquad+K_0(\alpha q)-K_0\left(\sqrt{1+\alpha^2}q\right)\Bigr\},
\label{sigma_exact_sinuous}
\end{eqnarray}
and for the peristaltic perturbation 
\begin{eqnarray}
F_p (q)&=&q^2-2 N_B\Bigl\{-\gamma_E +\ln\left({\alpha \over \sqrt{1+\alpha^2}}\right)
 - \ln(q/2) \nonumber \\
&&\ \ \  -K_0(q) -K_0(\alpha q)+K_0\left(\sqrt{1+\alpha^2}q\right)\Bigr\}.
\label{sigma_exact_varicose}
\end{eqnarray}

Let us now look at the limit of small $q$ for serpentine perturbations,
\begin{eqnarray}
F_s(q)&=&\left\{1-{N_B\over 4}\left[
\alpha^2\ln(1+\alpha^{-2})+\ln(1+\alpha^2)\right]\right\}q^2 \nonumber \\
&&-{ N_B\over 64}\Bigl\{(1+\alpha^2)^2\ln(1+\alpha^2)
-\alpha^4\ln(\alpha^2)\nonumber \\
&&\qquad ~~-6\alpha^2(1-{2\over 3}\gamma_E)\Bigr\} q^4\nonumber \\
&&-{ N_B\over 16}\alpha^2 q^4\ln({1\over 2} q)
 + {\cal O}\left(q^6,q^6\ln q\right)~.
\label{q_expand}
\end{eqnarray}

\dofloatfig
\fi

Serendipitously, the condition of stripe equilibrium is precisely
that which sets the coefficient of $q^2$ to zero.  This can be
interpreted as a consequence of rotational invariance.  
Note first that by assigning the same function $\zeta(y)$ to the bottom and
top edges of the stripe we have maintained the stripe width at $w$ to
linear order in $\zeta$, but not at quadratic order.  As shown in
Fig. \ref{stripe_pert_fig}, for a uniform tilt of the layer boundaries,
$\zeta_y={\rm constant}$, the width of the rotated stripe is
$w'=w/\sqrt{1+\zeta_y^2}\simeq w-(w/2)\zeta_y^2 + \cdots$.  Such
a uniform tilt will cost energy through the ``bulk" term
$E(w)$ in (\ref{stripe_energy1}) 
as $E(w')-E(w)\simeq -(w/2)E'(w)\zeta_y^2$, where
$E'\equiv dE/dw$.  
Now, the coefficient of $q^2$ in (\ref{q_expand}) has the interpretation of
an effective line tension, associated with an energy
\begin{equation}
E={1\over 2}\gamma_{\rm eff}\int\! dy \zeta_y^2~.
\label{rot_eqn}
\end{equation}
The rotational invariance argument thus shows that the
apparent surface tension vanishes at the equilibrium stripe width.
The surviving terms at ${\cal O}(q^4)$ look like bending energy of a rod,
\begin{equation}
E\sim {1\over 2}\int\! dy \zeta_{yy}^2,
\label{bend_eqn}
\end{equation}
but this interpretation is spoiled by the term $q^4\ln(q)$,
whose presence reflects the fundamental nonlocality of the
magnetic interactions.

\dofloatfig
\begin{figure}
\epsfxsize=3.3 truein
\centerline{\epsffile{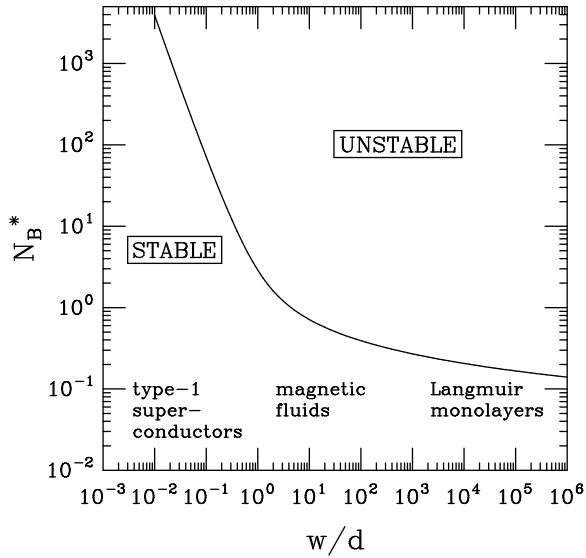}}
\smallskip
\caption[]{Critical Bond number for serpentine instability as a function
of stripe size.  Typical values of $w/d$ are indicated for three experimental
systems.}
\label{stripe_bond_fig}
\end{figure} 	
\fi

In the case of peristaltic perturbations, the force has a finite 
value as $q\to 0$,
reflecting the compressibility of the stripe.  The small-$q$ expansion is
\begin{eqnarray}
F_p &=& - 2 N_B\ln\left( {\alpha^2  \over 1+\alpha^2}\right) 
     + \Bigl\{ 1 - {N_B\over 2}\bigl[\alpha^2\ln\alpha \nonumber \\
 & &  - (1+\alpha^2)\ln\sqrt{1+\alpha^2}\bigr]\Bigr\} q^2 +
         {\cal O}\left(q^4,q^4\ln q\right) . 
\label{q_expand_varicose}
\end{eqnarray}

The equilibrium value of the stripe width as a function of
Bond number, deduced in (\ref{Bond_crit}), defines the boundary of
stability of stripes to serpentine perturbations.  Figure
\ref{stripe_bond_fig} displays this critical Bond number $ N_B^*$
as a function of $\alpha=w/d$.  At fixed $\alpha$, instability
occurs with increasing $N_B$, and likewise at fixed $N_B$ instability
sets in with increasing $\alpha$.  In the figure we have illustrated
the aspect ratios $\alpha$ corresponding not only to
type-I superconductors but also for typical experiments
on magnetic fluids in Hele-Shaw flow (with slab thicknesses and
stripe widths in the millimeter to centimeter range), and for
Langmuir monolayers (with domains up to tens of microns across and 
a {\it molecular} thickness to the layer).

At its equilibrium width, the energy of small distortions is
positive, vanishing as $q\to 0$.  Thus is would seem not possible
to find a field at which the stripe would be unstable to a finite-wavelength
mode.  This conclusion assumes that at any applied field the
stripe width has its equilibrium value (\ref{Bond_crit}).  
Fortuitously, the
elegant experimental observations on buckling instabilities
in Langmuir monolayers \cite{Stine} have shown us what happens
when this equilibrium is not reached.
Those observations concerned the
dynamics of buckling when the temperature was slowly increased.
Since these systems are near a critical 
point of phase separation, relatively small
changes in temperature produce large changes in the 
density difference between the phases (thus altering the 
discontinuity in dipole density $\Delta \mu$) and in the line tension.
These, of course, directly affect the stability of stripes, quantified
by the associated electric Bond number.  It was
observed that slow temperature ramps produced no buckling, while
rapid heating showed buckling.   This suggests \cite{Cebersreg} 
that the dependence of stability on ramp rate is associated with
a competition with mass transport as the stripe width adjusts
to keep up with the temperature. Under rapid ramps,
the width is out of equilibrium, yielding a nonzero (and potentially
destabilizing) coefficient of $q^2$.  Turning to the laminar state,
this suggests that in the early stages of flux penetration such a
mismatch between the actual and equilibrium widths allows the
buckling instability to occur.

\subsection{Elastic properties of the laminar state}

\begin{figure}
\epsfxsize=3.0 truein
\centerline{\epsffile{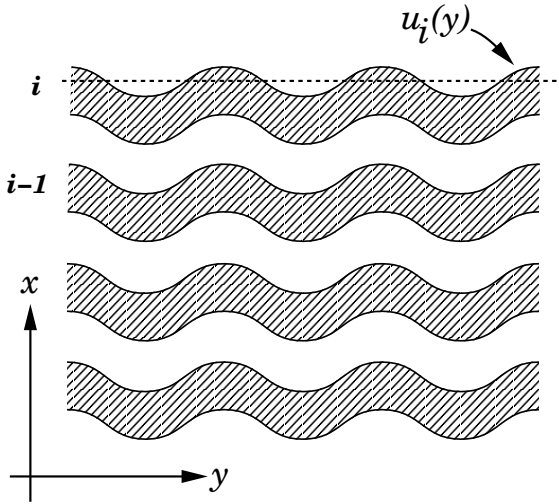}}
\smallskip
\caption[]{Schematic diagram showing the serpentine perturbations of the 
domain walls.  The displacement of the $i^{\rm th}$ domain wall is 
$u_{i}(y)$; in the continuum limit this will become the displacement 
field $u(x,y)$.}
\label{bend}
\end{figure}

By using the CL model we can also examine the elastic properties of the 
laminar state.  This is done by considering displacements $u_i(y)$ of the 
SN interfaces away from the equilibrium laminar phase, as shown in 
Fig.~\ref{bend}.  In the long wavelength (continuum) limit, $u_i(y)$
becomes the displacement field $u(x,y)$, and the effective elastic free energy becomes
\begin{equation}
{\cal F}_{\rm el} = \int\! d^2r
\left[ {B\over 2} \left(u_x + {1\over 2} u_{y}^2\right)^2 
+ {K_1 \over 2}  u_{yy}^2  \right], 
\label{elastic1}
\end{equation}
with $B$ the bulk (compressional) modulus and $K_1$ the bending modulus. 
This result applies to serpentine perturbations of the lamina; the 
peristaltic perturbations are gapped (like optical phonons), 
and do not contribute to the long wavelength properties. 
The general form of the free energy could have been anticipated from the 
single stripe calculations of the previous section; in particular, 
we see that the distortions in the $y$-direction appear as 
$u_{yy}^2$ (or $k_y^4 |u({\bf k})|^2$ in Fourier space), again signifying
that the effective surface tension is zero.  The nonlinear terms are 
required to preserve the rotational invariance of the free energy.  
The free energy, Eq.~(\ref{elastic1}),  is identical to the elastic free energy of 
a two dimensional smectic liquid crystal \cite{degennes}. This 
analogy is quite useful, as the properties of two dimensional 
smectics have been well studied; problems such as 
mechanical instabilities, thermal fluctuation effects, and defect 
structures have been considered.  We expect many of these same 
phenomena to occur in the laminar state.  

\dofloatfig
\fi

The elastic moduli for striped phases in two dimensional ferromagnets 
with dipolar interactions have been calculated in Refs.~\cite{kashuba93,ng95}, 
and for striped phases in ferrofluids in Ref.~\cite{cebers95,flament}. 
The calculation for the laminar phase is identical, and we include here only
the final results.  The bending modulus is  
\begin{eqnarray}
K_1 & = & {3 M^2 a^3 \over 8 \pi^4} \sum_{m=1}^{\infty} {\sin^2m\pi h \over m^5}
 \Biggl\{ 1 - \Biggl[1+\left({2\pi d m \over a}\right) \nonumber \\ 
& & \quad  + {1\over 3} \left({2\pi d m\over a}\right)^2 \Biggr] 
              e^{-2\pi d m/a} \Biggr\}, 
\label{bend_mod}
\end{eqnarray}
where the magnetization is $M= - H_c/4\pi$, and the equilibrium spacing 
$a=\sqrt{\Delta d /f_{\rm CL}}$, with $f_{\rm CL}(h,d,a)$ given by 
Eq.~(\ref{currentloop3}). In the thick film limit this becomes
\begin{equation}
K_1 = {3 M^2 a^3 \over 8 \pi^4} \sum_{m=1}^{\infty} {\sin^2m\pi h \over m^5}. 
\label{bend_mod1}
\end{equation}
The bulk modulus is
\begin{eqnarray}
B & = & a^2 \left[{\partial^2 (E_{CL}/A) \over \partial a^2}\right]_h \nonumber \\
  & = & {4\sigma_{SN} d \over a } - {8 M^2 d^2 \over a}\sum_{m=1}^{\infty}
{\sin^2 m\pi h \over m} e^{-2\pi d m/a} \nonumber \\
 & = & {4\sigma_{SN} d \over a } - {2 M^2 d^2 \over a}\ln\left[ 1 + 
   {\sin^2\pi h \over {\rm sinh}^2(\pi d/a)}\right]. 
\label{bulk}
\end{eqnarray}
In the thick film limit this becomes 
\begin{equation}
B = {4\sigma_{SN} d \over a }. 
\label{bulk2}
\end{equation}
The bending and bulk moduli may be combined to form the length scale 
$\tilde{\lambda} = \sqrt{K_1/B}$, which is a persistence
length for the distortion of the laminar structure
(not to be confused with the superconducting penetration depth).
For thick films this length becomes 
\begin{equation}
\tilde{\lambda}^2/a^2 = {3\over 32\pi^2} {\sum_{m=1}^{\infty} \sin^2(m\pi h)/m^{5}
  \over \sum_{m=1}^{\infty} \sin^2(m\pi h)/m^{3} } . 
\label{lambdatilde}
\end{equation}

\subsection{Dislocations in the laminar state} 
 
\begin{figure}
\epsfxsize=2.3 truein
\centerline{\epsffile{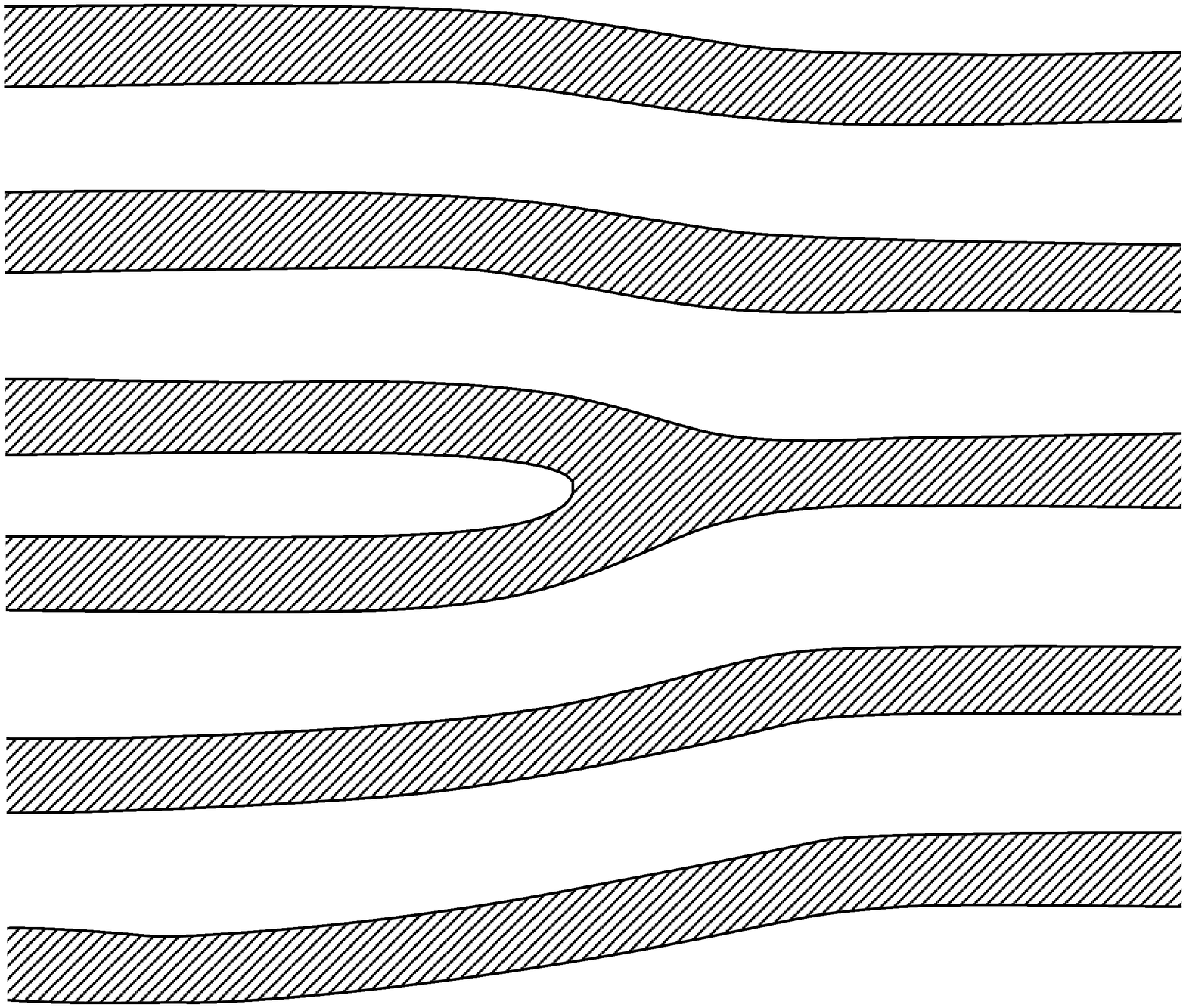}}
\smallskip
\caption[]{A dislocation in the laminar state with a Burger's vector of 1.}
\label{disloc}
\end{figure}

In many of the images of the laminar state \cite{Huebener} one often observes
edge dislocations, where half of a normal lamina has been inserted into the 
laminar structure (see Fig.~\ref{disloc} for a schematic diagram).  
\dofloatfig
\fi
By using 
our elastic theory, we can determine the displacement field produced by such a 
dislocation; this problem has been studied in the context of 2D smectics 
\cite{toner81} and our calculation closely follows Ref.~\cite{toner81}. 
We begin with the linearized Euler-Lagrange equations for the defect displacement 
field $u^D(x,y)$, 
\begin{equation}
u_{xx}^D - \tilde{\lambda}^2 u_{yyyy}^D = m a \delta'(x)\theta(y), 
\label{disloc1}
\end{equation}
where a source term has been added to the right hand side to account for the 
presence of the dislocation, in such a way that the line integral of
$\nabla u$ around the dislocation is $m a$ (the Burger's vector), 
with $m$ the number of half sheets inserted and $a$ the lamina spacing. 
Equation (\ref{disloc1}) may be solved using Fourier transforms \cite{toner81}, 
with the result that 
\begin{equation}
u^{D}(x,y) = {m a \over 4} {\rm sgn}(x) \left[ {\rm erf}
\left({y\over \sqrt{4\tilde{\lambda} |x|}}\right) + 1 \right]. 
\label{disloc2}
\end{equation} 
For an collection of dislocations centered at $\{ {\bf r}_i\}$ with 
strengths $\{m_i\}$, we may introduce the dislocation density 
\begin{equation}
m({\bf r}) = \sum_i m_i \, \delta( {\bf r} - {\bf r}_i), 
\label{disloc3}
\end{equation}
so that the displacement field is obtained by linear superposition,
\begin{equation}
u^D({\bf r}) = \int d^2r'\, m({\bf r}') G({\bf r} - {\bf r}'),
\label{disloc4}
\end{equation}
with 
\begin{equation}
G({\bf r} - {\bf r}') = {a\over 4} {\rm sgn}(x-x') \left[ {\rm erf}
\left( { y - y'\over\sqrt{4\tilde{\lambda}|x-x'|}}\right) + 1 \right].
\label{disloc5}
\end{equation}
The effective free energy for the defects, ${\cal F}_D$, is then obtained by substituting
Eq.~(\ref{disloc4}) into the the elastic free energy, and using the 
``harmonic conjugate'' trick of Toner and Nelson \cite{toner81}.  The final result
is  
\begin{eqnarray}
{\cal F}_D &=& {1\over 2} \int d^2r_1\, \int_{|{\bf r}_1 - {\bf r}_2|>a}  d^2r_2\,
 m({\bf r}_1) m({\bf r}_2) U({\bf r}_1 - {\bf r}_2) \nonumber \\
 & & \ \ + E_D \int d^2r\, m^2({\bf r}),
\label{disloc6}
\end{eqnarray}
where the interaction potential is 
\begin{equation}
U({\bf r}) = {a^2 B\over 4} \left(\tilde{\lambda} \over \pi |x|\right)^{1/2}
               e^{-y^2/4\tilde{\lambda} |x|},
\label{disloc7}
\end{equation}
and the defect core energy is
\begin{eqnarray}
E_D & =& {B\over 2} \int d^2r\, \left[ \tilde{\lambda}^2\left( G_{yy}\right)^2
                       + (G_x)^2 \right] \nonumber \\
   & = & {1\over 8\sqrt{\pi}} B a^2 \left({\tilde{\lambda}\over a}\right)^{1/2}. 
\label{core_energy}
\end{eqnarray}
The core energy can be calculated as a function of the reduced field $h$ by using
the results of the previous section [Eqs.~(\ref{bulk}) and (\ref{lambdatilde})].  
A rough estimate shows that 
this energy is generally of order $10^{-3} (H_c^2/8\pi) a^3$, and can therefore
be quite small; as a result it should be easy to nucleate dislocations in the 
laminar phase.

\section{Discussion}

The free-boundary approach to the energetics and dynamics of the
intermediate state has led us to a clearer understanding of the 
shape instabilities of individual flux domains and ordered arrays. 
At the same time, the correspondence with smectic liquid crystals 
suggests that phenomena observed there should have an analog
in type-I superconductors.  Below we suggest several experiments
to visualize these phenomena. These have as their
starting point the ordered laminar state, produced with an
in-plane component to the field. 
Preliminary experiments by Reisin and Lipson \cite{Reisin} have shown
some of these phenomena.

\noindent (i) {\it The buckling instability}:
Rapid changes in the magnitude of the applied normal field
may allow buckling instabilities to occur in much the same way
as observed in Langmuir monolayers.  Of interest would be
the dependence of buckling wavelength on the magnitude of
the field jump.

\noindent (ii) {\it The chevron instability}:
If the in-plane magnetic field is applied at an angle with
respect to the lamina we expect an instability toward a
zig-zag or chevron pattern as the lamina attempt to reorient.
This is the analog of the Helfrich-Hurault effect in smectics
\cite{degennes}, wherein a field component normal to smectic layers, 
producing a torque on 
them, induces an undulatory instability.

\noindent (iii) {\it The Eckhaus instability}:
If the magnetic field normal to the slab is slowly increased or
decreased in magnitude the stripe width and spacing must adjust to stay
in equilibrium.  As in convective systems \cite{CrossHohenberg}, 
this may occur through an Eckhaus-like nucleation phenomenon to create or
destroy laminae. Dislocations can be produced which will
move toward the sample edges or annihilate at the center
in accord with the direction in which the wavelength must adjust.
Their climb and glide dynamics will provide an important testing
ground for the theory of superconductor-normal interface dynamics. 

\noindent (iv) {\it Critical-point effects}:
In the simplified contour dynamics in which the local field
is taken to be the critical field we saw that the effective
Bond number depended only on the ratio $d/\Delta(T)$.  Near
the zero-field critical temperature for the superconductor-metal
transition the interfacial width $\Delta(T)$ diverges
with reduced temperature $(T_c-T)/T_c$.
This should produce characteristic changes in the equilibrium stripe
width as well as possibly inducing shape transformations.

We close by emphasizing what has {\it not} been accomplished in this
study.  First, we have considered laterally infinite samples, so that
the whole issue of flux penetration at the edges is ignored.
This is known to be very significant in both type-I and type-II
superconductors \cite{Duran}.  A treatment of these effects requires
not only the electromagnetics of the fields 
in the neighborhood of the slab edges but also consideration
of processes such as domain fission.  Second, we have presented
a highly oversimplified dynamical picture in which diffusional
instabilities are absent and only mechanical ones appear.
The interplay between the Mullins-Sekerka and these mechanical
instabilities appears not to have been considered theoretically
and may shed some light on various problems in flux domain pattern
formation.  Third, a large-scale numerical study of the many-domain
problem has not been attempted, precluding a clear understanding of
the true ``energy landscape" of this strongly interacting system. 
Fourth, the effects of in-plane components to the applied magnetic
fields has not been incorporated into the free-boundary approach
in any quantitative way.  This will be important for a quantitative
understanding of the instabilities described above.
Fifth, the extension of matched asymptotic methods used in
purely two-dimensional systems to the slab geometry in which 
the intermediate state appears has not been developed.  A
detailed study of this point would greatly clarify the free-boundary
approach to flux domain shapes.  Finally, coarse-graining approaches to 
domain dynamics analogous to Otto and Kohn's recent study of magnetic fluid 
pattern formation may prove quite fruitful \cite{Otto,OttoKohn}.

\acknowledgments

We are grateful to C.-Y. Mou, Thomas R. Powers, and John Toner for 
important discussions on the energetics of the intermediate state, 
to O. Narayan and H. Bokil for communication of 
Ref. \protect{\cite{Narayan}} prior to publication, and
to C. Reisin and S.G. Lipson for sharing with us the results of their
ongoing experiments.
This work was supported in part by NSF PFF Grant 
No. DMR 93-50227 (REG),  No. DMR 92-23586 and 96-28926 (ATD), and the Alfred P. Sloan
Foundation (REG and ATD).

\appendix
\section{Flux-stripe stability calculation}

Here we collect some details of the stability analysis for single
flux stripes.  
In the case of peristaltic distortions, 
\begin{equation}
{\bf r}_{\pm}(y)=y\hat{\bf e}_y
+\left(\pm\frac{w}{2}\pm\zeta(y)\right)\hat{\bf e}_x,
\label{app1}
\end{equation}
The linearized normal force can be reduced to the form
\begin{eqnarray}
U(y)&=&\gamma \zeta_{yy}\nonumber \\
&&+2M^2\int_{-\infty}^{\infty}\! dy'
\left[\zeta(y')+\zeta(y)\right]\left[S_{d^2+w^2}-S_{w^2}\right]\nonumber \\
&&+2M^2\int_{-\infty}^{\infty}\! dy'
\left[\zeta(y')-\zeta(y)\right]\left[S_{d^2}-S_{0}\right]~,
\label{app2}
\end{eqnarray}
where $S_{a^2}=1/\vert(y'-y)^2+a^2\vert$.
For serpentine distortions, the displacements are
\begin{equation}
{\bf r}_{\pm}(y)=y\hat{\bf e}_y
+\left(\pm\frac{w}{2}+\zeta(y)\right)\hat{\bf e}_x~,
\label{app3}
\end{equation}
and the normal force has the simpler form
\begin{eqnarray}
U(y)&=&\gamma \zeta_{yy}\nonumber \\
&&+2M^2\int_{-\infty}^{\infty}\! dy'
\left[\zeta(y')-\zeta(y)\right]\nonumber \\
&&\qquad\qquad \times \left[S_{d^2+w^2}-S_{w^2}-S_{d^2}-S_{0}\right]~,
\label{app4}
\end{eqnarray}
If the distortion is the plane wave $\zeta(y)=A\cos(ky)$, then the force
is proportional to $\zeta$ with a coefficient $F(k)$
\begin{eqnarray}
F(k)&=& -\gamma k^2+4M^2\!\!\int_0^{\infty}\! dy 
\left[1-\cos\left(ky\right)\right]\nonumber \\
&&\qquad\qquad\qquad\times \left[S_{d^2+w^2}-S_{w^2}-S_{d^2}+S_{0}\right] 
\label{sigmafirst}
\end{eqnarray}
where now $S_{a^2}=1/\vert y^2+a^2\vert$.
Several of the integrals with integrands proportional to $\cos(ky)$ 
reduce trivially to Bessel functions, but care must
be taken to account for canceling divergences in the remaining terms.  This is
conveniently done by considering the limiting process 
\begin{equation}
I=\lim_{u\to\infty}\int_0^u\! dy 
\left\{S_{d^2+w^2}-S_{w^2}-S_{d^2}
+{1-\cos(ky)\over y} \right\}~.
\label{I1}
\end{equation}
Rescaling these equations, we obtain
\begin{eqnarray}
I&=&\lim_{u\to\infty}\bigg\{ 
\int_0^{u/\sqrt{w^2+h^2}}\! dy S_1
-\int_0^{u/w}\! dy S_1\nonumber \\
&&\qquad-\int_0^{u/h}\! dy S_1
+\int_0^{uk}\! dy{1-\cos(y)\over y} \bigg\}\nonumber \\
&& \label{I2}
\end{eqnarray}
A useful intermediate result at this stage is \cite{GR}
\begin{equation}
\int_0^u\! dy {1-\cos(y)\over y}=\gamma_E +\ln(u) 
+ \int_u^{\infty}\! dy {\cos(y)\over y},
\label{intermed}
\end{equation}
where $\gamma_E =0.577215\ldots$ is Euler's constant. 
Substitution into
(\ref{I2}) and evaluation of the remaining integrals then yields
the final results (\ref{sigma_exact_sinuous}) and
(\ref{sigma_exact_varicose}) for the energy of
serpentine and peristaltic perturbations.

\vfill\eject

\widetext

\begin{table}
\caption{Analogies between free streamline flow in fluids and lamina formation
         in the intermediate state of type-I superconductors.}
\label{table1}
\begin{tabular}{ll}
Free streamline flow around a plate & Laminae in superconductors \\
\tableline

Complex potential $w = \phi + i \psi$  & Complex potential $w = \phi + i A_y$\\

Complex fluid velocity $ u - i v = - d w / d \zeta $  
            & Complex magnetic field $B = B_x - i B_z = - d w / d \zeta$ \\

Streamlines                     & Field lines (lines of force)  \\

Free streamline                 & Superconducting-normal interface \\

Free streamline velocity $U$    & Superconducting critical field $H_{c}$ \\

Region of fluid flow            & Normal phase with nonzero magnetic field \\

Cavity behind plate             & Superconducting phase  \\

Riabouchinsky flow              & Lamina in a finite thickness plate 

\end{tabular}
\end{table}

\vfil
\eject
\rlap{\phantom{.}}
\vfil
\eject

\begin{table}
\caption{Analogies between interfacial energetics of type-I superconductors and
other systems.  The energy of a set $\{{\cal D}_i\}$ of domains is
written as
$
E[\{{\bf r}_i\}]=\Pi \sum_i A_i + \gamma\sum_i L_i -{1\over 2}\Omega
\oint\! ds \oint\! ds' \hat{\bf t}_i\cdot \hat{\bf t}_j 
\Phi_{ij}(R_{ij}/\xi)~.
$
}
\label{table2}
\begin{tabular}{llllll}

System & $\Pi$ & $\gamma$ & $\Omega$ & $\Phi$ & $\xi$ \\
\tableline

type-I superconductors$^a$ & $(H_c^2d/8\pi)(\rho_n+h^2/\rho_n)$ & $H_c^2d\Delta/8\pi$ & 
$H_a^2d/8\pi^2$ & $\sinh^{-1}(1/z)+z-\sqrt{1+z^2}$ & $d$ \\

magnetic fluids$^b$ & Lagrange multiplier & $d\sigma_{FW}$ & 
$2dM^2$ & $\sinh^{-1}(1/z)+z-\sqrt{1+z^2}$ & $d$ \\

Langmuir monolayers$^c$ & Lagrange multiplier & $\gamma_{LE-LC}$ & 
$(\Delta \mu)^2$ & $1/2z^*$ & $d_{\rm mol}$ \\

FitzHugh-Nagumo model$^e$& $\Delta F$ & $\bar D$ & 
$\rho$ & $K_0(z)$ & $1^{**}$

\end{tabular}
\noindent {\it Explanation of symbols:} $\sigma_{FW}$, ferrofluid water surface
tension; $M$, ferrofluid magnetization; $\gamma_{LE-LC}$, line tension between
liquid expanded (LE) and liquid condensed (LC) phases in a Langmuir monolayer;
$\Delta\mu$, discontinuity in electric dipole moment density between 
LE and LC phases; $d_{\rm mol}$, a molecular cutoff -- monolayer thickness.

\noindent $^a$ Present work.

\noindent $^b$ Refs. \protect{\cite{pra,pre}}. 

\noindent $^c$ Refs. \protect{\cite{LeeMcConnell,jpc}}. 

\noindent $^d$ Eq. (\ref{FN}) and Refs. \protect{\cite{turingprl,turingpre}}. 

\noindent $^*$ This limiting form is supplemented with a cutoff procedure.
See Ref. \protect{\cite{jpc}}.

\noindent $^{**}$ The system of units in Eq. (\ref{FN}) sets the inhibitor
screening length to unity.

\end{table}


\begin{references}

\bibitem[*]{emaild} Electronic mail: dorsey@phys.ufl.edu

\bibitem[**]{emailr} Electronic mail: gold@physics.arizona.edu.

\bibitem{Huebener} R.P. Huebener, {\it Magnetic Flux Structures in
Superconductors} (Springer-Verlag, New York, 1979).

\bibitem{Faber} T.E. Faber, Proc. Roy. Soc. A {\bf 248}, 460 (1958).

\bibitem{Haenssler} F. Haenssler and L. Rinderer, Helv. Phys.
Acta {\bf 40}, 659 (1967).

\bibitem{Goren} R.N. Goren and M. Tinkham,
J. Low Temp. Phys. {\bf 5}, 465 (1971).

\bibitem{magnetooptic} R.P. Huebener, R.T. Kampwirth, and V.A. Rowe,
Cryogenics {\bf 12}, 100 (1972).

\bibitem{Sharvin} I.V. Sharvin, Sov. Phys. JETP {\bf 6}, 1031 (1958).

\bibitem{Landau} L.D. Landau, Sov. Phys. JETP {\bf 7}, 371 (1937).

\bibitem{others} E.R. Andrew, Proc.\ Roy.\ Soc.\ (London)
{\bf A194}, 98 (1948); K. Maki, Ann. Phys. {\bf 34}, 363 (1965);
G. Lasher,  Phys. Rev. {\bf 154}, 345 (1967);  D.J.E. Callaway, Ann. Phys. {\bf
213}, 166 (1992).

\bibitem{Dorsey91} H. Frahm, S. Ullah, and A.T. Dorsey, Phys. Rev.
Lett. {\bf 66}, 3067 (1991). 

\bibitem{Liu91} F. Liu, M. Mondello, and N. Goldenfeld,
Phys. Rev. Lett. {\bf 66}, 3071 (1991). 

\bibitem{Pippard} A.B. Pippard, Phil. Mag. {\bf 41}, 243 (1950).

\bibitem{Mullins} W.W. Mullins and R.F. Sekerka, J. Appl. Phys.
{\bf 35}, 444 (1964).

\bibitem{asymptotics} A.T. Dorsey, Ann. Phys. {\bf 233}, 248
(1994); J.C. Osborn and A.T. Dorsey, Phys. Rev. B {\bf 50}, 15 961 (1994);
S. J. Chapman, Quart. Appl. Math. {\bf 53}, 601 (1995).

\bibitem{Pearl} J. Pearl, Appl. Phys. Lett. {\bf 5}, 65 (1964). 

\bibitem{Fetter} A.L. Fetter and P.C. Hohenberg, Phys. Rev. {\bf 159}, 
          330 (1967).

\bibitem{goldstein96} R.E. Goldstein, D.P. Jackson, and A.T. Dorsey,
Phys. Rev. Lett. {\bf 76}, 3818 (1996).

\bibitem{Kesslerrev} For a review, see D.A. Kessler, J. Koplik, and H. Levine,
Adv. Phys. {\bf 37}, 255 (1988).

\bibitem{Narayan} See H. Bokil and O. Narayan, 
``Flux penetration in slab shaped Type-I superconductors," cond-mat/9610039.

\bibitem{Seulscience} M. Seul and D. Andelman,
Science {\bf 267}, 476 (1995).

\bibitem{Rosensweig} R.E. Rosensweig, M. Zahn, and 
R. Shumovich, J. Magn. Magn. Mater. {\bf 39}, 127 (1983).

\bibitem{cebers} A.O. Tsebers and M.M. Mairov, Magnetohydrodynamics
{\bf 16}, 21 (1980).

\bibitem{science} A.J. Dickstein, S. Erramilli, R.E. Goldstein,
D.P. Jackson, and S.A. Langer, Science {\bf 261}, 1012 (1993).

\bibitem{pra} S.A. Langer, R.E. Goldstein, and D.P. Jackson,
Phys. Rev. A {\bf 46}, 4894 (1992).

\bibitem{pre} D.P. Jackson, R.E. Goldstein, and
A.O. Cebers, Phys. Rev. E {\bf 50}, 298 (1994).

\bibitem{Andelman} D. Andelman, F. Brochard, and J.-F. Joanny, J.
Chem. Phys. {\bf 86}, 3673 (1987).

\bibitem{LeeMcConnell} K.Y.C. Lee and H.M. McConnell, J. Phys. Chem.
{\bf 97}, 9532 (1993).

\bibitem{jpc} R.E. Goldstein and D.P. Jackson, J. Phys. Chem. {\bf 98}, 9626 (1994).

\bibitem{Seulfilms} M. Seul, L.R. Monar, L.O'Gorman, and R. Wolfe,
Science {\bf 254}, 1616 (1991).

\bibitem{turingprl} D.M. Petrich and R.E. Goldstein, Phys. Rev. Lett.
{\bf 72}, 1120 (1994).

\bibitem{turingpre} R.E. Goldstein, D.J. Muraki, and D.M. Petrich, Phys.
Rev. E {\bf 53}, 3933 (1996).

\bibitem{Hagberg} A. Hagberg and E. Meron, Phys. Rev. Lett. {\bf 72},
2494 (1994).

\bibitem{Muratov} C.B. Muratov and V.V. Osipov, Phys. Rev. E {\bf 53},
3101 (1996).

\bibitem{Lee} K.J. Lee, W.D. McCormick, Q. Ouyang, and H.L. Swinney,
Science {\bf 261}, 192 (1993).

\bibitem{Leelong} K.J. Lee and H.L. Swinney, Phys. Rev. E {\bf 51},
1899 (1995).

\bibitem{FitzHugh} R. FitzHugh, Biophys. J. {\bf 1}, 445 (1961); 
J.S. Nagumo, S. Arimoto, and Y. Yoshizawa, Proc. IRE {\bf 50}, 2061 (1962); 
R. FitzHugh, in {\it Biological Engineering}, ed.  H.P. Schwan 
(McGraw-Hill, New York, 1969).

\bibitem{milnethomson} L. M. Milne-Thomson, {\it Theoretical Hydrodynamics, 
Fifth Edition} (MacMillan Press Ltd., London 1968), Chapters XI and XII. 

\bibitem{birkhoff} G. Birkhoff and E. H. Zarantonello,
{\it Jets, Wakes, and Cavities} (Academic Press, New York 1957). 

\bibitem{LandauLifshitz} L.D. Landau, E.M. Lifshitz, and L.P.
Pitaevskii, {\it Electrodynamics of Continuous Media}, 2nd ed.
(Pergamon Press, New York 1984), p. 119.

\bibitem{fortini} A. Fortini and E. Paumier, 
Phys. Rev. B {\bf 5}, 1850 (1972).

\bibitem{magenergy} The energy of a sample of volume $V$ with a 
uniform magnetization ${\bf M}$ in an external field ${\bf H}_a$ is 
$-{1\over 2} {\bf M}\cdot {\bf H}_a V$. For a superconductor,
${\bf M} = - {\bf H}_n/4\pi $, so that the energy per unit area 
of the sample is $H_n H_a d/4\pi$. 

\bibitem{tension} This ignores the added area of the SN interface
due to the flaring of the normal domains. 

\bibitem{Yu} C.-Y. Mou, private communication.

\bibitem {kooy60} C. Kooy and V. Enz, Philips Res. Rep. {\bf 15},
7 (1960).

\bibitem{cape71} J. A. Cape and G. W. Lehman, J. App. Phys. {\bf 42},
5732 (1971).

\bibitem{abrikosovbook}  See A.A. Abrikosov, {\it Fundamentals of the Theory
of Metals} (North-Holland, Amsterdam 1988), p. 479.

\bibitem{dendrite}  For a discussion of surface tension anisotropy
within the context of dendritic growth, see Ref.~\cite{Kesslerrev}. 

\bibitem{bodmer} A. Bodmer, U. Essman, H. Tr\"auble, Phys. Stat.
Solidi (a) {\bf 13}, 471 (1972); A. Bodmer, {\it ibid.} {\bf 19},
513 (1973).

\bibitem{Stine} K.J. Stine, C.M. Knobler, and R.C. Desai, Phys.
Rev. Lett. {\bf 65}, 1004 (1990).

\bibitem{Cebersreg} A.O. Cebers and R.E. Goldstein, unpublished.

\bibitem{degennes} W. Helfrich, Appl. Phys. Lett. {\bf 17}, 531 (1970);
J.P. Hurault, J. Chem. Phys. {\bf 59}, 2086 (1973);  See also
P.G. de Gennes and J. Prost, {\it The Physics of Liquid Crystals} 
(Oxford University Press, New York 1993), pp. 361--4.

\bibitem{kashuba93} A. Kashuba and V.L. Pokrovsky, Phys. Rev. Lett.
{\bf 70}, 3155 (1993); Phys. Rev. B {\bf 48}, 10 355 (1993).
Some errors in the calculation of the bending modulus are 
corrected in Appendix A of Ar. Abanov, V. Kalatsky, V.L. Pokrovsky, 
and W.M. Saslow, Phys. Rev. B {\bf 51}, 1023 (1995). 

\bibitem{ng95} K.-O. Ng and D. Vanderbilt, Phys. Rev. B {\bf 52}, 2177 (1995).

\bibitem{cebers95} A. Cebers, J. Magn. Mag. Mat. {\bf 149}, 93 (1995).

\bibitem{flament} C. Flament, J.C. Bacri, A. Cebers, F. Elias, R. Perzynski,
Europhys. Lett. {\bf 34}, 225 (1996). 

\bibitem{toner81} J. Toner and D. R. Nelson, Phys. Rev. B {\bf 23}, 316 (1981).

\bibitem{Reisin} C.R. Reisin and S.G. Lipson, ``Period Dependence of the 
Intermediate State Structures of Type-I Superconductors," preprint (1997),
and private communication.

\bibitem{CrossHohenberg} M.C. Cross and P.C. Hohenberg, Rev. Mod. Phys.
{\bf 65}, 851 (1993).

\bibitem{Duran} C.A. Duran, P.L. Gammel, R.E. Miller, and D.J. Bishop,
 Phys. Rev. B {\bf 52}, 75 (1995).

\bibitem{Otto} F. Otto, ``Dynamics of labyrinthine pattern formation in magnetic
fluids: a mean field theory,'' Arch. Rat. Mech. Analysis, in press (1997).

\bibitem{OttoKohn} F. Otto and R. Kohn, ``Small surface energy, 
coarse-graining, and selection of microstructure,'' Physica D (in press). 

\bibitem{GR} I.S. Gradshteyn and I.M. Ryzhik, {\it Table of Integrals, 
Series, and Products, Fifth Edition} (Academic Press, Inc., San Diego 1994), 
\# 3.782.1. 

\end{references}
\end{document}